\documentclass[twocolumn,useAMS,usenatbib]{mnras} \usepackage{natbib}

\usepackage{amsmath,latexsym,amssymb} \usepackage{epsfig,graphicx} \topmargin-1cm
\usepackage{epstopdf}
\usepackage{float}
\usepackage{fixltx2e}
\usepackage{color}
\usepackage{url}
\graphicspath{{./draft4f_figs/}}
\DeclareGraphicsExtensions{.pdf,.png,.jpg,.eps}

\def\reff@jnl#1{{\rm#1\/}}

\def\aj{\reff@jnl{AJ}}                  
\def\araa{\reff@jnl{ARA\&A}}            
\def\apj{\reff@jnl{ApJ}}                        
\def\apjl{\reff@jnl{ApJ}}               
\def\apjs{\reff@jnl{ApJS}}              
\def\apss{\reff@jnl{Ap\&SS}}            
\def\aap{\reff@jnl{A\&A}}               
\def\aapr{\reff@jnl{A\&A~Rev.}}         
\def\aaps{\reff@jnl{A\&AS}}             
\def\baas{\reff@jnl{BAAS}}              
\def\jcap{\reff@jnl{JCAP}}              
\def\jrasc{\reff@jnl{JRASC}}            
\def\memras{\reff@jnl{MmRAS}}           
\def\mnras{\reff@jnl{MNRAS}}            
\def\physrep{\reff@jnl{Phys.Rep.}}
\def\pra{\reff@jnl{Phys.Rev.A}}         
\def\prb{\reff@jnl{Phys.Rev.B}}         
\def\prc{\reff@jnl{Phys.Rev.C}}         
\def\prd{\reff@jnl{Phys.Rev.D}}         
\def\prl{\reff@jnl{Phys.Rev.Lett}}      
\def\pasp{\reff@jnl{PASP}}              
\def\pasj{\reff@jnl{PASJ}}              
\def\skytel{\reff@jnl{S\&T}}            
\def\solphys{\reff@jnl{Solar~Phys.}}    
\def\sovast{\reff@jnl{Soviet~Ast.}}     
\def\ssr{\reff@jnl{Space~Sci.Rev.}}     
\def\nat{\reff@jnl{Nature}}             

\newcommand{\hmpc}{\ensuremath{h^{-1}\mathrm{Mpc}}}
\newcommand{\hkpc}{\ensuremath{h^{-1}\mathrm{kpc}}}
\newcommand{\hMsun}{\ensuremath{h^{-1}M_{\odot}}}

\newcommand{\beq}{\begin{equation}}
\newcommand{\eeq}{\end{equation}}
\newcommand{\beqa}{\begin{eqnarray}}
\newcommand{\eeqa}{\end{eqnarray}}

\title[Intrinsic Alignments]{Intrinsic alignments of disk and elliptical galaxies in the MassiveBlack-II and Illustris simulations}
\author[Tenneti et al.]
{Ananth Tenneti$^1$\thanks{\tt vat@andrew.cmu.edu},
Rachel Mandelbaum$^1$\thanks{\tt rmandelb@andrew.cmu.edu},
Tiziana Di Matteo$^1$\thanks{\tt tiziana@phys.cmu.edu}
\\$^1$McWilliams Center for Cosmology, Department of Physics, Carnegie Mellon University, Pittsburgh, PA 15213, USA
 }

\date{\today}
\begin{document}
\maketitle

\begin{abstract}
We study the shapes and intrinsic alignments of disks and elliptical galaxies in the MassiveBlack-II
(MBII) and
Illustris cosmological hydrodynamic simulations, with volumes of $(100\hmpc)^{3}$ and
$(75\hmpc)^{3}$ respectively. We find that simulated disk galaxies are more oblate in shape and more
misaligned with the shape of their host dark matter subhalo when compared with ellipticals. The disk
major axis is found to be oriented towards the location of nearby elliptical galaxies. We also find
that the disks are thinner in MBII and misalignments with dark matter halo orientations are smaller
in both disks and ellipticals when compared with Illustris. As a result, the intrinsic alignment
correlation functions at fixed mass have a higher amplitude in MBII than in Illustris. Finally, at scales above $\sim 0.1$\hmpc, the intrinsic alignment two-point correlation functions for disk galaxies in both
simulations are consistent with a null detection, unlike those for ellipticals.
Despite significant differences in the treatments of hydrodynamics and
baryonic physics in the simulations, we find that the $w_{\delta +}$ correlation function scales similarly with transverse separation. However, the less massive galaxies show different scale dependence in the ED correlation. This result indicates that, while hydrodynamic simulations are a promising tool to study intrinsic alignments, further study is needed to understand the impact of differences in the implementations of hydrodynamics and baryonic feedback.
\end{abstract}


\section{Introduction} \label{S:intro}
\vspace{1in}
Weak lensing is a promising cosmological probe that can help in understanding the nature of dark matter, dark energy and modified theories of gravity \citep{{2006astro.ph..9591A},2013PhR...530...87W}. Future weak lensing surveys such as the Large Synoptic Survey Telescope\footnote{\url{http://www.lsst.org/lsst/}}
(LSST; \citealt{LSST09}),
Euclid\footnote{\url{http://sci.esa.int/euclid/},
  \url{http://www.euclid-ec.org}} \citep{LAA+11}, and the Wide-Field 
Infrared Survey Telescope\footnote{\url{http://wfirst.gsfc.nasa.gov}}
(WFIRST; \citealt{SGB+15}) should constrain cosmological parameters such as the dark energy equation
of state to sub-percent levels. However, the intrinsic alignment of galaxies, the coherent
correlations of the galaxy shapes with each other and with the underlying density field, is a
significant astrophysical systematic in weak lensing analysis
\citep{{2000MNRAS.319..649H},{2000ApJ...545..561C},{2002MNRAS.335L..89J},{2004PhRvD..70f3526H}}. Ignoring
the effects of intrinsic alignments on a weak lensing analysis can bias the estimation of the dark
energy equation of state parameter. \citet{2015arXiv150608730K} find that without marginalizing over
intrinsic alignments, this bias in the value of $w$ can be up to $\sim 80 \%$ of its value. Hence, an
understanding of intrinsic alignments and their scaling with galaxy mass, luminosity, redshift and
morphological type is necessary to develop effective mitigation strategies. Further, studies of intrinsic alignments can also help in understanding the physics of galaxy formation and evolution \citep{2012PhRvD..86h3513S,2013JCAP...12..029C,2015arXiv150602671S}. For reviews of intrinsic alignments, see \cite{2014arXiv1407.6990T},
\cite{2015arXiv150405456J}, \cite{2015arXiv150405465K}, and \cite{2015arXiv150405546K}.       

Intrinsic alignments have been studied analytically using the linear alignment model
\citep{2004PhRvD..70f3526H}, extensions that include nonlinear contributions
\citep{{2007NJPh....9..444B},{2015arXiv150402510B}} and the halo model
\citep{2010MNRAS.402.2127S}. $N$-body simulations have also been used to study IA by stochastically
populating halos with galaxies with a random orientation or by using semi-analytic methods
\citep{2006MNRAS.371..750H,2013MNRAS.436..819J}. Recently, hydrodynamic simulations of cosmological
volumes, such as the MassiveBlack-II \citep{2015MNRAS.450.1349K}, Horizon-AGN
\citep{2014MNRAS.444.1453D}, EAGLE \citep{2015MNRAS.446..521S} and Illustris
\citep{{2014Natur.509..177V},{2014MNRAS.444.1518V},{2014MNRAS.445..175G}} simulations have emerged
as a useful tool to study intrinsic alignments. They enable direct predictions of intrinsic
alignments of the stellar component of galaxies using a large statistical sample, including the
physics of galaxy formation
\citep{{2014MNRAS.441..470T},{2015MNRAS.448.3522T},{2015MNRAS.448.3391C},{2015arXiv150404025V},{2015arXiv150706996V},{2015arXiv150707843C}}. In
previous work \citep{2014MNRAS.441..470T}, we studied the shapes of the stellar component of the
galaxies using the MassiveBlack-II cosmological hydrodynamic simulation and compared our results
with observational measurements finding good agreement. In a follow up study
\citep{2015MNRAS.448.3522T}, we  measured the two-point correlations of the galaxy shapes with the
density field and found that the scaling of the correlation function measured in the simulations is
consistent with observational results and, on large scales, with predictions of the tidal alignment model. 

However, none of these studies have considered morphological divisions of the
galaxy sample into disks and ellipticals.  There are theoretical and observational motivations for such a split. The
galaxies  for which intrinsic alignments have been robustly
measured in real data
\citep{{2006MNRAS.367..611M},{2007MNRAS.381.1197H},{2009ApJ...694..214O},{2011A&A...527A..26J},{2013MNRAS.432.2433H},{2015MNRAS.450.2195S}} 
are predominantly elliptical
galaxies, for which alignments on large scales are well described by the linear alignment model
\citep{2011JCAP...05..010B}. Observationally, there has
been no significant detection of intrinsic alignment shape correlations for disk galaxies
\citep{{2007MNRAS.381.1197H},{2011MNRAS.410..844M}}, except for a hint of a detection for
the most luminous blue sample in \cite{{2007MNRAS.381.1197H}} at low significance. Due to the importance of angular momentum in the formation of disk
galaxies, their intrinsic alignments are likely 
described by the quadratic alignment model, for which the shape-density correlation
vanishes in the case of a Gaussian density field \citep{2004PhRvD..70f3526H}. However, in general, due to the non-linear evolution of the density field, we expect a non-zero correlation.  

Given the
different mechanisms for the alignments of elliptical and disk galaxies, it will be interesting to
investigate these differences in a large-volume hydrodynamic simulation simulation. The alignments of disk galaxies have been studied previously using small volume hydrodynamic
simulations
\citep[e.g.,][]{{2005ApJ...627L..17B},{2010MNRAS.404.1137B},{2010MNRAS.405..274H},{2011MNRAS.415.2607D}}. The
most recent hydrodynamic simulations with cosmological volumes have a resolution high enough to
enable dynamical classification of galaxies into  disks and ellipticals using a method 
described in \cite{2003ApJ...597...21A}. Recently, the galaxies in the
Illustris simulation (based on a moving mesh code) have been found to have
a disk galaxy fraction that compares favorably with observations
\citep{2014MNRAS.444.1518V}, so it will be interesting to study the intrinsic alignments of disk
galaxies using this simulation. Further, it has been shown 
that disk galaxies in simulations based on SPH and moving mesh techniques differ in
properties such as disk scale lengths and angular momentum \citep{2012MNRAS.427.2224T}. Since
MassiveBlack-II is an SPH-based hydrodynamic simulation, we can similarly explore differences in
properties such as the disk galaxy fraction, specific angular momentum and alignments. In this
paper, we first compare the intrinsic alignments of galaxies in MBII and Illustris for galaxies in a
similar stellar mass range. This comparison will show how 
different implementations of hydrodynamics or baryonic physics, and box size effects, can affect
predictions of galaxy intrinsic alignments. We then
compare the alignments for galaxy subsamples that have been kinematically classified  into disks and ellipticals. 

This paper is organized as follows. In Section~\ref{S:Simulations}, we describe the details of the
MassiveBlack-II and Illustris simulations used in the study. In Section~\ref{S:Methods}, we describe
the methods adopted to measure the shapes of galaxies, quantify intrinsic alignments, and
kinematically classify galaxies. In Section~\ref{S:Results}, we compare the galaxy shapes in MBII
and Illustris and their two-point correlations in a similar stellar mass range. In
Section~\ref{S:morph}, we show the results for  galaxy shapes and their intrinsic alignments
separately for disks and elliptical galaxies in both simulations. Finally, a summary of our conclusions is given in Section~\ref{S:conclusions}. 
             
\section{Simulations}\label{S:Simulations}
In this study, we use the MassiveBlack-II (MB-II) hydrodynamic simulation and publicly-released data from the  Illustris
simulation \citep{2015arXiv150400362N}.

\subsection{MassiveBlack-II Simulation}
 MB-II is a state-of-the-art high
resolution, large volume, cosmological hydrodynamic simulation of
structure formation. This simulation has been performed with {\sc
  p-gadget}, which is a hybrid version of the parallel code, {\sc
  gadget2} \citep{2005MNRAS.361..776S} upgraded to run on Petaflop
scale supercomputers. In addition to gravity and smoothed-particle
hydrodynamics (SPH), the {\sc
  p-gadget} code also includes the physics of multiphase ISM model
with star formation \citep{2003MNRAS.339..289S}, black hole accretion
and feedback
\citep{2005MNRAS.361..776S,2012ApJ...745L..29D}. Radiative cooling and
heating processes are included \citep[as in][]{1996ApJS..105...19K},
as is photoheating due to an imposed ionizing UV background. The black hole accretion and feedback are modeled according to \cite{2005Natur.433..604D} based on quasar-mode feedback. Here, a fixed fraction ($5\%$) of the radiative energy release by the accreted gas is assumed to couple thermally to the nearby gas and this is independent of the accretion rate. The
details of this simulation can be found in \cite{2015MNRAS.450.1349K}. 

MB-II contains $N_\mathrm{part} = 2\times 1792^{3}$ dark matter and gas
particles in a cubic periodic box of length $100$\hmpc\ on a side,
with a gravitational smoothing length $\epsilon = 1.85$\hkpc\ in
comoving units. A single dark matter particle has a mass $m_\text{DM} =
1.1\times 10^{7}\hMsun$ and the initial mass of a gas particle is
$m_\text{gas} = 2.2\times 10^{6}\hMsun$, with the mass of each star
particle being $m_\text{star} = 1.1\times 10^{6}\hMsun$. The cosmological
parameters used in the simulation are as follows: amplitude of matter
fluctuations $\sigma_{8} = 0.816$, spectral index $\eta_{s} = 0.96$,
mass density parameter $\Omega_{m} = 0.275$, cosmological constant
density parameter $\Omega_{\Lambda} = 0.725$, baryon density parameter
$\Omega_{b} = 0.046$, and Hubble parameter $h = 0.702$ as per WMAP7
\citep{2011ApJS..192...18K}.

\subsection{Illustris Simulation}
The Illustris simulation is performed with the AREPO TreePM-moving-mesh code
\citep{2010MNRAS.401..791S}  in a box of volume $(75\hmpc)^3$. The simulation follows $1820^3$ dark
matter particles and an approximately equal number of baryonic elements with a gravitational
smoothing length of $1.4$ comoving kpc for the dark matter particles. The mass of each dark matter
particle is $4.41 \times 10^{6}\hMsun$ and the initial baryonic mass resolution is $8.87
  \times 10^{5}\hMsun$. The galaxy formation physics includes subgrid-model for star formation and
associated supernova feedback, black hole accretion and feedback. Here, the black hole
  accretion and feedback are modeled according to quasar-mode feedback at high accretion rates  and
  radio-mode feedback at low accretion rates. In the radio-mode feedback, it is assumed that the
  accretion periodically produces an AGN jet that inflates hot bubbles in the surrounding gas. When
  the black hole has increased its mass by a certain fraction, an AGN-driven bubble is created. The
  accretion rate and current black hole mass determine the duty cycle of bubble injection, energy
  content and radius of bubbles. This model is different from that of MBII where the radio-mode feedback is absent and the quasar-mode feedback is independent of accretion rate. A detailed description of the models adopted in Illustris can be found in \cite{2013MNRAS.436.3031V}. The cosmological
parameters are as follows: $\sigma_{8} = 0.809$, $\eta_{s} = 0.963$, $\Omega_{m} = 0.2726$, $\Omega_{\Lambda} = 0.7274$, $\Omega_{b} = 0.0456$, $h = 0.704$.  

\subsection{Galaxy and halo catalogs}
In both simulations, halo catalogs are generated using
the friends of friends (FoF) halo finder algorithm
\citep{1985ApJ...292..371D}. The FoF algorithm identifies halos on the
fly using a linking length of $0.2$ times the mean interparticle
separation. The subhalo catalogs are generated using the {\sc
  subfind} code \citep{2001MNRAS.328..726S} on the halo catalogs. The
subhalos are defined as locally overdense, self-bound particle
groups. In this paper, we analyze the shapes of the stellar components in the subhalos and their two-point correlation functions. Based on convergence tests in
\cite{2014MNRAS.441..470T}, we only analyze the measured stellar shapes if there
are $\ge 1000$ dark matter and star particles. As a result, we exclude $\sim
10$\% ($1$\%) of the galaxies in MBII (Illustris) in the $10^{9}$--$10^{10}\hMsun$ stellar mass bin due
to their subhalo masses being low enough that they do not have 1000 dark matter particles.  This
fraction reduces to $<1 \%$ in the stellar mass bin from $10^{10}$--$10^{11}\hMsun$, while no
galaxies are discarded for  $M*>10^{11}\hMsun$.

\section{Methods}\label{S:Methods}
\begin{figure}
\begin{center}
\includegraphics[width=3.2in]{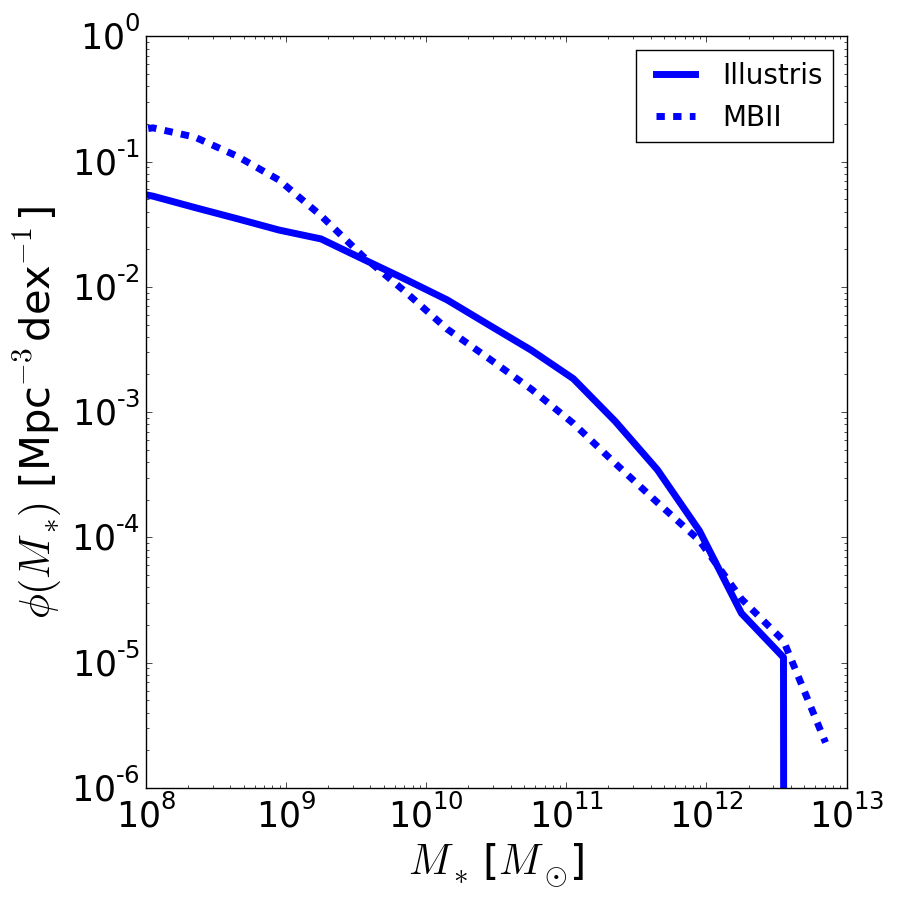}
\caption{\label{F:fig1smf} Comparison of galaxy stellar mass functions in MBII and Illustris at
  $z=0.06$. 
}
\end{center}
\end{figure}
Here we describe the methods used to measure galaxy shapes, quantify intrinsic alignments, and
kinematically classify the galaxies into disk and elliptical samples. 

\subsection{Galaxy shapes}\label{SS:shapedef}
The shapes of the stellar matter component in subhalos are modeled as
ellipsoids in three dimensions using 
the eigenvalues and eigenvectors of the iterative version of the reduced inertia tensor given by :
\begin{equation} \label{eq:redinertensor}
\widetilde{I}_{ij} = \frac{\sum_{n} m_{n}\frac{x_{ni}x_{nj}}{r_{n}^{2}}}{\sum_{n} m_{n}}
\end{equation}
where the summation is over particles indexed by $n$, and 
\begin{equation} \label{eq:rn2}
 r_{n}^{2} = \frac{x_{n0}^{2}}{a^{2}} + \frac{x_{n1}^{2}}{b^{2}} + \frac{x_{n2}^{2}}{c^{2}}.
\end{equation}
where $a,b,c$ are half-lengths of the principal axes of the ellipsoid and are all equal to $1$ in the first iteration. The reduced inertia tensor gives more weight to particles that are
closer to the center, which reduces the influence of loosely bound particles present in the outer
regions of the subhalo. Additionally, this method corresponds more closely to observational shape measurements such as the ones based on weighted quadrupole moments (see \citealt{2015arXiv150405465K}) where more weight is given to particles in the inner regions.  
The eigenvectors of the inertia tensor are
${\hat{e}_{a}, \hat{e}_{b}, \hat{e}_{c}}$ with corresponding
eigenvalues 
$\lambda_{a} > \lambda_{b} > \lambda_{c}$. The eigenvectors represent
the principal axes of the ellipsoid, with the half-lengths of the principal
axes $(a,b,c)$ given by 
$(\sqrt{\lambda_{a}},\sqrt{\lambda_{b}},\sqrt{\lambda_{c}})$. The 3D
axis ratios are 
\begin{equation} \label{eq:axisratios}
q = \frac{b}{a}, \,\, s = \frac{c}{a}.
\end{equation}

The
projected shapes are calculated by projecting the positions of the particles
onto the $XY$ plane and modeling the shapes as ellipses. 


We note here that without using the iterative scheme, the reduced
  inertia tensor will lead to shape estimates that are biased to rounder values due to the spherical
  symmetry imposed by the $1/r^{2}$ weighting. This has been discussed in \cite{2015MNRAS.448.3522T}; a detailed description of the iterative procedure and further details regarding other definitions of the inertia tensor used to calculate shapes and their impact on intrinsic alignments can also be found there.

\subsection{Misalignment angle}\label{SS:ma}
To study the relative orientation between the shapes defined by the 
dark matter and stellar matter components in subhalos, we compute 
the probability distributions of misalignment angles as in \cite{2014MNRAS.441..470T}. 
If $\hat{e}_{da}$ and $\hat{e}_{ga}$ are the major axes of the shapes
defined by the dark matter and stellar matter components,
respectively, then we 
define the misalignment angle by
\begin{equation} \label{eq:misalignangle}
 \theta_{m} = \arccos(\left|\hat{e}_{da} \cdot \hat{e}_{ga}\right|).
\end{equation}

\subsection{Two-point statistics}\label{SS:pes}

\begin{figure*}
\begin{center}
\includegraphics[width=3.2in]{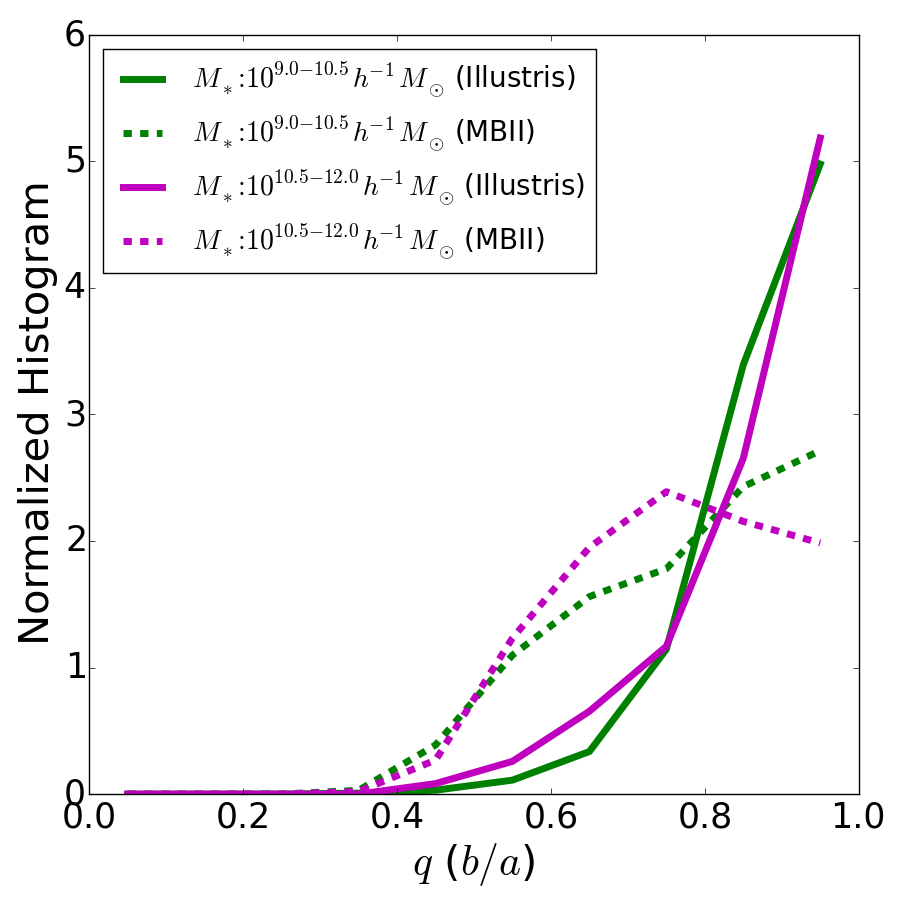}
\includegraphics[width=3.2in]{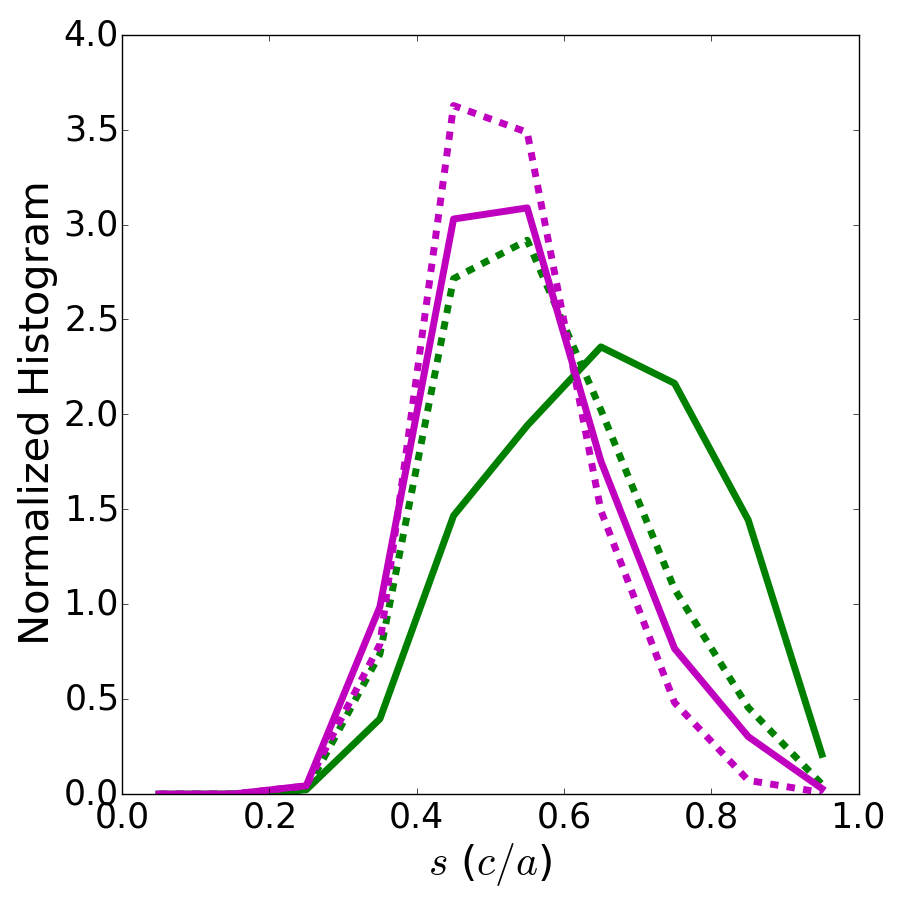}
\caption{\label{F:fig1_qs} Normalized histograms of axis ratios (left: $q$, right:
  $s$) of the 3D shapes of galaxies in MBII and Illustris in two stellar mass bins:
  $10^{10.0-11.0}h^{-1}M_{\odot}$ and $10^{11.0-12.0}h^{-1}M_{\odot}$.
}
\end{center}
\end{figure*}

The intrinsic alignments of galaxies with the large-scale
density field are quantified using the ellipticity-direction (ED) and the
projected shape-density ($w_{\delta +}$) correlation functions. 
The ED correlation quantifies the position angle alignments of galaxies
in 3D, while the projected shape correlation function can be used to compare
against observational measurements that include the 2D shape of the galaxy.

The ED correlation function cross-correlates the orientation of the
major axes of subhalos with the large-scale density
field. For a subhalo centered at position \textbf{x} with major axis
direction $\hat{e}_{a}$, let the unit vector in the direction of a
tracer of the 
matter density field at a distance $r$ be $\hat{\textbf{r}} =
\textbf{r}/r$. Following the notation of \cite{2008MNRAS.389.1266L},
the ED cross-correlation function is given by
\begin{equation} \label{eq:ED3d}
 \omega(r) = \langle \mid \hat{e}_{a}(\textbf{x})\cdot \hat{\textbf{r}}(\textbf{x}) \mid^{2} \rangle - \frac{1}{3}
\end{equation}
which is zero for galaxies randomly oriented according to a uniform distribution.

The matter density field can be represented using either the positions
of dark matter particles (in which case the correlation function is
denoted by the symbol $\omega_{\delta}$) or the positions of subhalos
(in which case it is simply
denoted $\omega$). Here, we only use $\omega_{\delta}$ to eliminate
the effect of subhalo bias.

The projected shape correlation functions are computed to directly
compare our results from simulations with observations. Here, we
follow the notation of \cite{2006MNRAS.367..611M} to define 
the galaxy-intrinsic shear correlation function,
$\hat{\xi}_{g+}(r_{p},\Pi)$ and the corresponding projected
two-point statistic, $w_{\delta +}$. Here, $r_{p}$ is the comoving
transverse separation of a pair of galaxies in the $XY$ plane and
$\Pi$ is their separation along the $Z$ direction.

The projected shape correlation function, 
$w_{\delta +}(r_{p})$ is 
given by
\begin{equation} \label{eq:wgp}
 w_{\delta +}(r_{p}) =
 \int_{-\Pi_\text{max}}^{+\Pi_\text{max}}\hat{\xi}_{\delta +}(r_{p},\Pi)\,\mathrm{d}\Pi
\end{equation}
We calculated the correlation functions over the whole length
of the box, $L_{box}$ with $\Pi_\text{max} = L_{box}/2$, where the length of
  the box is $100\hmpc$ and $75 \hmpc$ in MBII and Illustris respectively.  The details
  regarding the calculation of $\hat{\xi}_{\delta +}(r_{p},\Pi)$ using the projected shapes and
  density field traced by dark matter particles can be found in \cite{2015MNRAS.448.3522T}. The projected
correlation functions are obtained via direct summation. The error bars for the ED and
  $w_{\delta +}$ correlation functions are calculated using the jackknife method, where the
  correlation function for each jackknife sample is calculated by eliminating one eighth of the volume of the box.

\subsection{Bulge-to-disk decomposition}\label{S:morphbd}
In order to identify a galaxy according to its morphological type, we follow the procedure from \cite{2009MNRAS.396..696S} and define a circularity parameter for each star within $10$ times the stellar half-mass radius, $\epsilon = j_{z}/j_\text{circ}(r)$. Here $j_{z}$ is the component of the specific angular momentum of the star in the direction of the total angular momentum of the galaxy calculated using all star particles within $10$ times the stellar half-mass radius. $j_\text{circ}(r)$ is the specific angular momentum of a circular orbit at the same radius as the star,
\begin{equation} \label{eq:jcirc}
 j_\text{circ}(r) = rV_\text{circ}(r) = \sqrt{\frac{GM(<r)}{r}}.
\end{equation}

All stars with $\epsilon > 0.7$ are identified as disk stars. We
then define the bulge-to-total ratio as $BTR = 1 - f_{\epsilon >0.7}$, where $f_{\epsilon >0.7}$ is
the fraction of stars belonging to the disk. In this paper, we classify the galaxies
with $BTR < 0.7$ as disk galaxies and the galaxies with $BTR$ greater than this value as elliptical
galaxies.  However, we will briefly explore the results of varying this threshold in Sec.~\ref{SS:diskfrac}.
\section{Galaxy shapes and alignments in Illustris and MBII}\label{S:Results}
\subsection{Galaxy stellar mass function}

Before examining the galaxy shape distributions and alignments in the two simulations, we first
present some background about the simulated galaxy samples.

In Figure~\ref{F:fig1smf}, we compare the galaxy stellar mass function in MBII and
Illustris at $z=0.06$. At lower masses, the density of galaxies is higher in MBII, while at higher
masses the two simulations  are similar. \cite{2015MNRAS.450.1349K} compare the galaxy stellar mass function in MBII with
observations, noting that MBII overpredicts
the mass function at $z=0.06$ at both low and high mass. The lower mass over-prediction can be
resolved by only considering galaxies with a non-zero star-formation rate, which suggests a need for
a star formation and stellar feedback model with an associated mass dependent
wind \citep[e.g.,][]{{2008MNRAS.387..577O},{2010MNRAS.406..208O}}
\cite{2014MNRAS.444.1518V} discuss
the stellar mass function of Illustris simulation in greater detail,  where they also report a
higher galaxy density at the fainter end compared with observations.  As shown, both MBII and
Illustris contain a reasonably-sized galaxy population for the stellar mass range
$10^{9.0-12.0}\hMsun$. For the rest of this paper, we use this stellar mass range, which is also
consistent with our convergence criterion ($\ge 1000$ star particles). 

\subsection{Shapes and misalignment angles}\label{SS:shapes-mass}

\begin{figure}
\begin{center}
\includegraphics[width=3.2in]{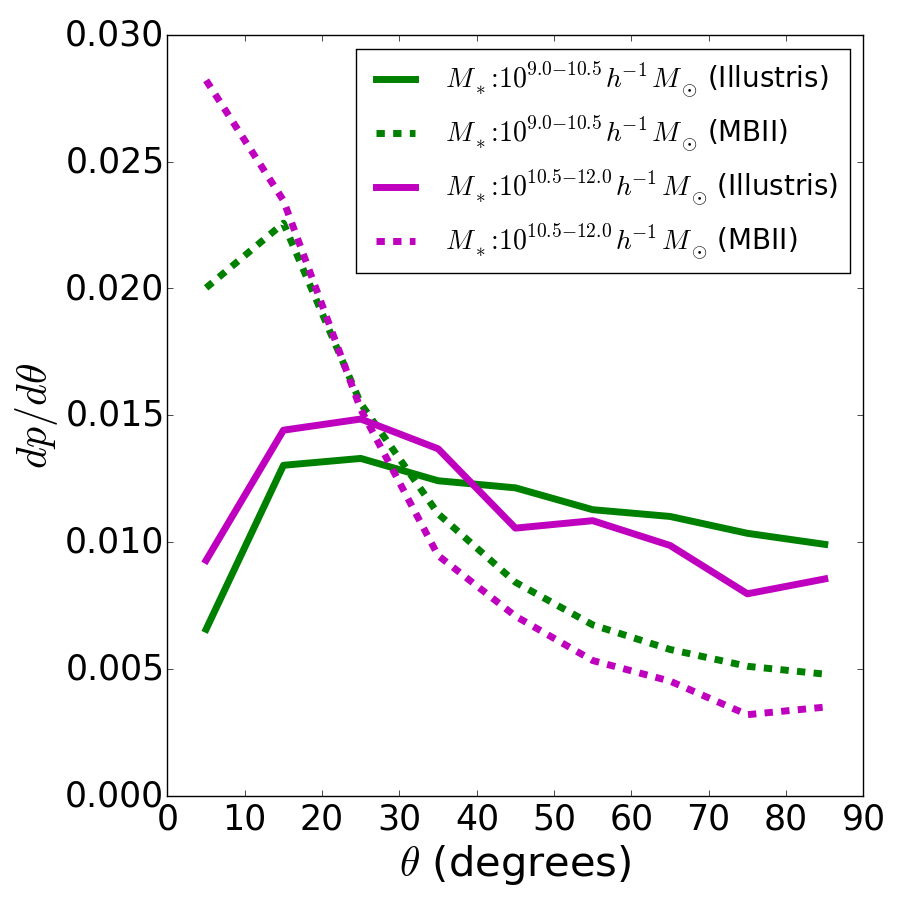}
\caption{\label{F:fig1_ma}Normalized histograms of the misalignment angles between 3D shapes of
  galaxies and their host dark matter subhalos 
  in MBII and Illustris in two stellar mass bins: $10^{9.0-10.5}h^{-1}M_{\odot}$ and
  $10^{10.5-12.0}h^{-1}M_{\odot}$.}
\end{center}
\end{figure}

In this section, we explore the differences in the shapes and orientations of galaxies in MBII and Illustris, focusing on the mass dependence without including a morphological classification.
In Fig.~\ref{F:fig1_qs}, we compare the distribution of axis ratios $q~(b/a)$ and $s~(c/a)$ in
Illustris and MBII in two stellar mass bins, $10^{9.0-10.5}h^{-1}M_{\odot}$ and
$10^{10.5-12.0}h^{-1}M_{\odot}$. Both axis ratios ($q$ and $s$) are larger in
Illustris, which means that within the same broad mass range, galaxy shapes are rounder compared to MBII. Further,
the shapes are rounder in galaxies of lower mass, consistent with results presented in
\cite{2014MNRAS.441..470T}. Table~\ref{T:tab12} shows the mean axis ratios, $\langle q \rangle$ and
$\langle s \rangle$, within the two stellar mass bins. The mean values of $q$ in
Illustris are larger by a factor of $\sim 12-14 \%$ when compared with MBII, while the mean values of
$s$ differ by $\sim 4-11 \%$. 

We compare the normalized histograms of 3D misalignment angles
(Eq.~\ref{eq:misalignangle}) between the orientations of the galaxy shape and the corresponding dark
matter subhalo in Figure~\ref{F:fig1_ma}. The mass dependence of the misalignment angles is similar
in Illustris and MBII, with the galaxies being more misaligned in the lower mass range. However,
the galaxy shapes are significantly more misaligned with their host dark matter subhalos in the
Illustris simulation, with the mean misalignment angles differing by $\sim 50-60 \%$ when compared with MBII. The mean misalignment angles in the two stellar mass bins are given in Table~\ref{T:tab3}. 

\begin{table}
\begin{center}
\caption{\label{T:tab12} Mean axis ratios, $\langle q \rangle$ and $\langle s \rangle$, 
  for galaxies in Illustris and MBII. 
}
\begin{tabular}{@{}lcccc}
\hline
 & \multicolumn{2}{c}{Illustris} & \multicolumn{2}{c}{MBII}\\
\hline
 $M_*$ (\hMsun) & $\langle q \rangle$ & $\langle s \rangle$ & $\langle q \rangle$ & $\langle s \rangle$ \\
\hline
 $10^{9.0}-10^{10.5}$ & $0.88$ & $0.65$ & $0.78$ & $0.56$\\
 $10^{10.5}-10^{12.0}$ & $0.87$ & $0.54$ & $0.76$ & $0.52$\\
\end{tabular}
\end{center}
\end{table}

\begin{table}
\begin{center}
\caption{\label{T:tab3} Mean 3D misalignment angles, $\langle \theta \rangle$ (degrees), between the
  major axis of galaxies and their host dark matter subhalos in Illustris and MBII.}
\begin{tabular}{@{}lcccc}
\hline
 $M_*$ (\hMsun) & Illustris & MBII\\
\hline
$10^{9.0}-10^{10.5}$ & $45.04 \pm 0.16^{\circ}$ & $31.33 \pm 0.11^{\circ}$ \\
$10^{10.5}-10^{12.0}$ & $41.51 \pm 0.54^{\circ}$ & $26.49 \pm 0.46^{\circ}$ \\
\end{tabular}
\end{center}
\end{table}

\subsection{Two-point statistics}

\begin{figure*}
\begin{center}
\includegraphics[width=3.2in]{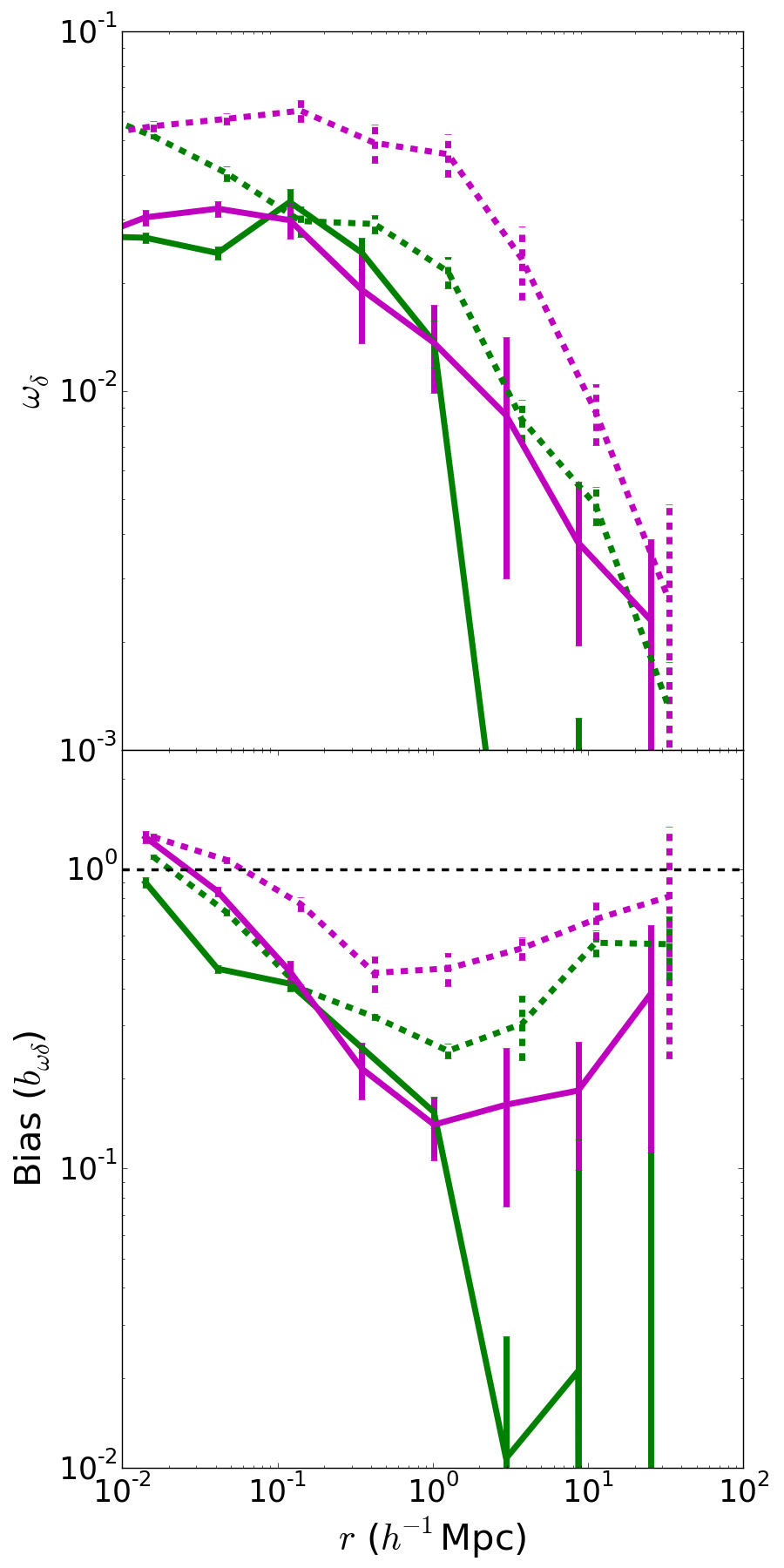}
\includegraphics[width=3.2in]{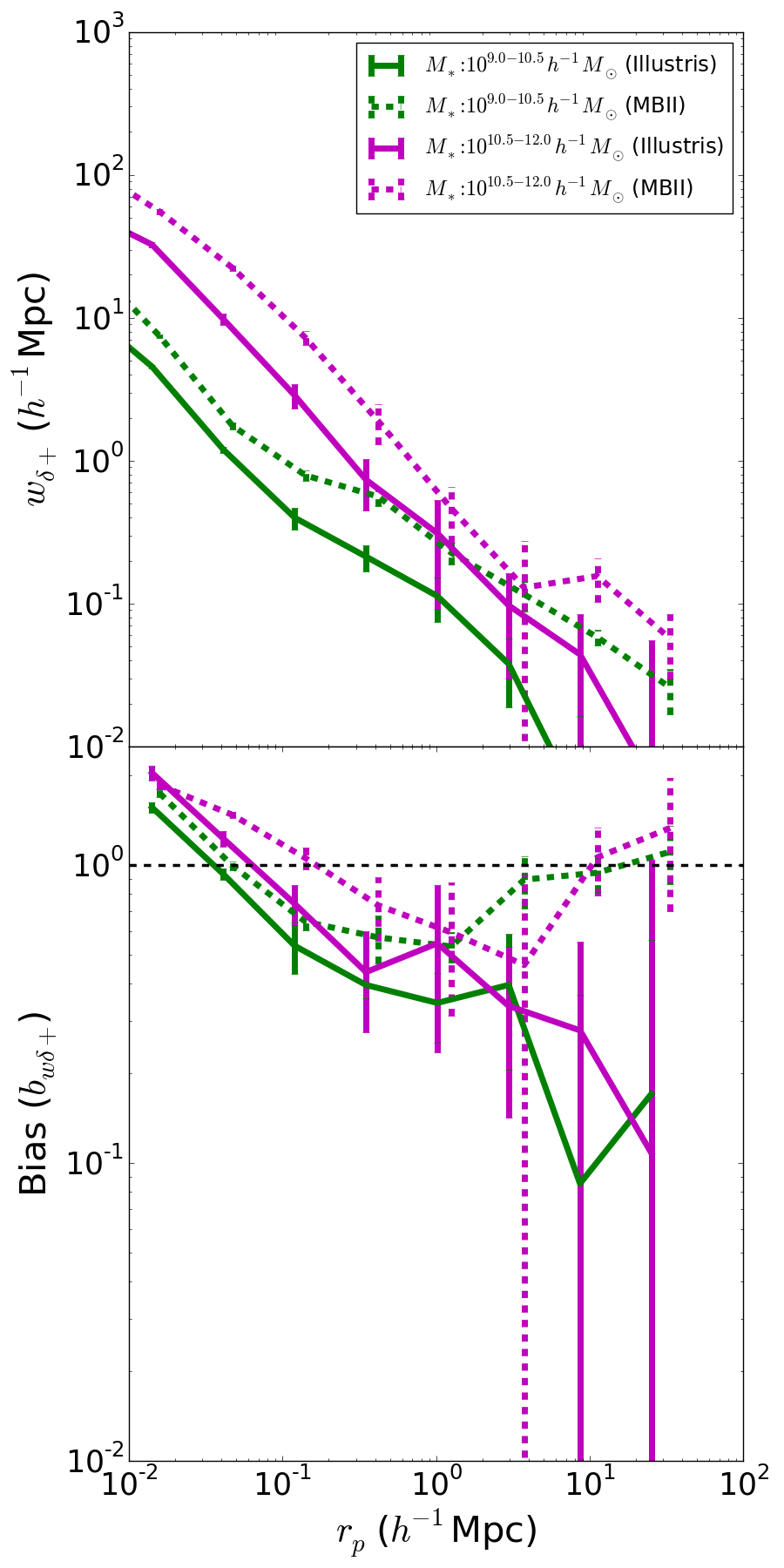}
\caption{\label{F:fig1_edwdp} {\em Top:} Ellipticity-direction (ED) and projected shape-density ($w_{\delta
    +}$) correlation functions of the stellar components of galaxies in MBII and Illustris in two stellar
  mass bins, $10^{9.0-10.5}h^{-1}M_{\odot}$ and $10^{10.5-12.0}h^{-1}M_{\odot}$. {\em Left:} ED; {\em
    Right:} $w_{\delta +}$. {\em Bottom}: the biases, $b_{\omega \delta}$ and $b_{w\delta +}$,
  defined as the ratios of the ED and $w_{\delta +}$ correlations functions of stellar components to
  the same correlation function computed using the dark matter subhalo.
}
\end{center}
\end{figure*}

\begin{figure*}
\begin{center}
\includegraphics[width=3.2in]{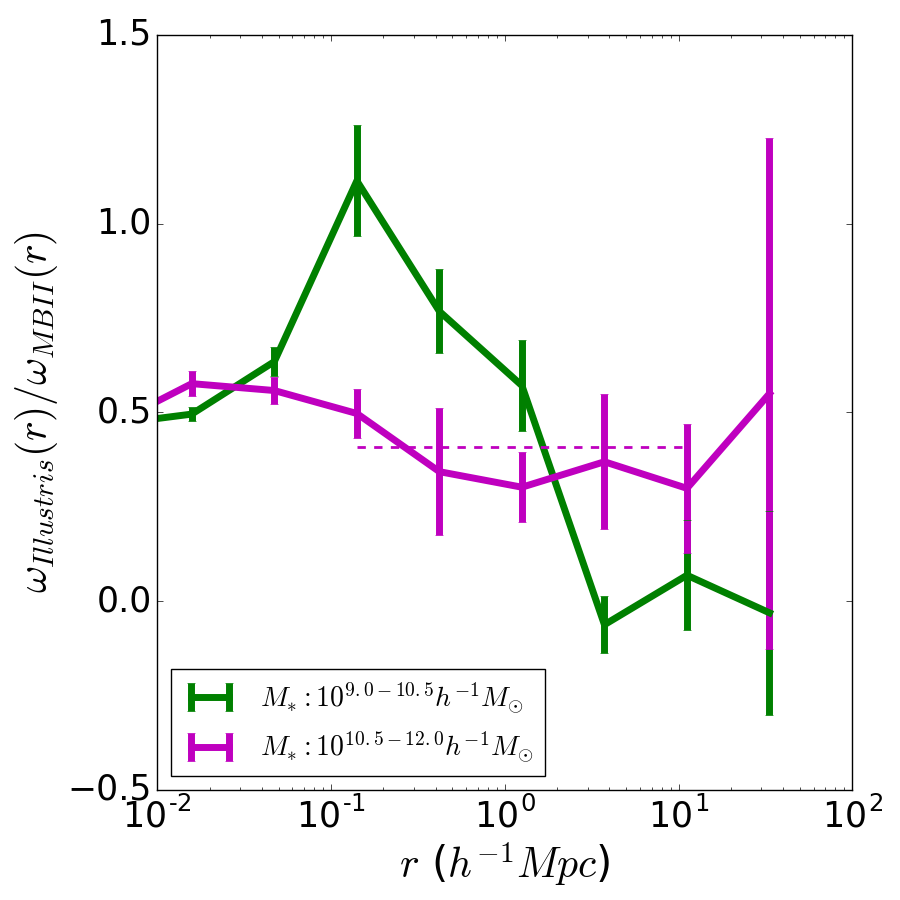}
\includegraphics[width=3.2in]{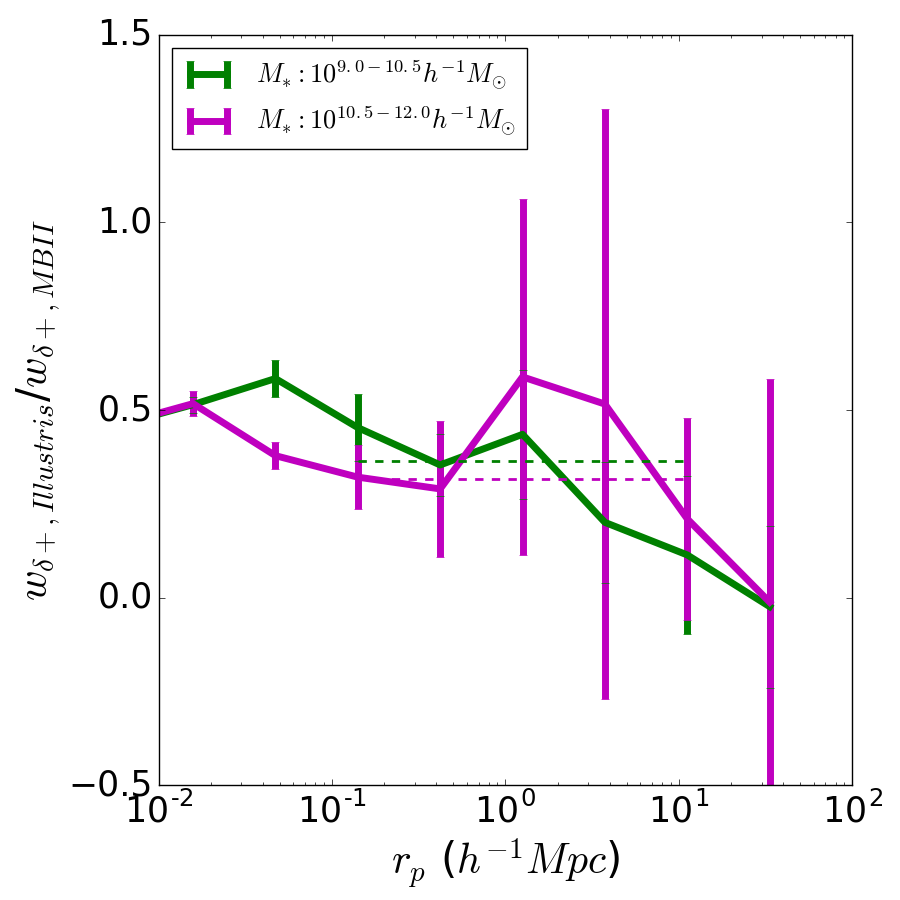}
\caption{\label{F:fig1_edwdpratio} Ratio of 
ellipticity-direction (ED, {\em (left)}) and projected shape-density ($w_{\delta
    +}$, {\em (right)}) correlation functions of the stellar components of galaxies in MBII and Illustris in two stellar
  mass bins, $10^{9.0-10.5}h^{-1}M_{\odot}$ and $10^{10.5-12.0}h^{-1}M_{\odot}$. 
The horizontal lines represent the best fit values obtained after fitting the ratio to a straight line in the range $0.1-10\hmpc$ 
and the line color indicates the corresponding mass bin.
}
\end{center}
\end{figure*}

The correlation of galaxy shapes with the density field can be quantified using the two-point
statistics, ED and $w_{\delta +}$, defined in Section~\ref{SS:pes}. In the top left panel of 
Figure~\ref{F:fig1_edwdp}, we compare the ED correlation functions in Illustris and MBII in two 
stellar mass bins, $10^{9.0-10.5}h^{-1}M_{\odot}$ and $10^{10.5-12.0}h^{-1}M_{\odot}$. 
 Due to the larger misalignment of the stellar shapes in Illustris with their
host dark matter subhalos (shown in Fig.~\ref{F:fig1_ma}), the amplitude of these two-point
correlation functions is lower than in MBII for the same stellar mass range. 
The radial
  scaling of the ED correlation function is similar for the two 
simulations in the higher mass bin
  and increases with mass in MBII. 
Given that the differences in the mean misalignment angles of
  Illustris are smaller in between the two mass bins when compared with MBII, 
the mass dependence of
  the ED correlation function in Illustris is less significant. 
In the top right panel of
Figure~\ref{F:fig1_edwdp}, we compare the $w_{\delta +}$ correlation functions in the same stellar
mass bins. The amplitude of $w_{\delta +}$ is smaller in 
Illustris than in MBII for two reasons: the larger misalignment of the stellar shapes with the dark
matter subhalos, and fact that 
the galaxy shapes are rounder in Illustris. However, the radial scaling and mass dependence is
similar to that of MBII. To further understand and quantify the radial scaling of the
  correlation functions in the two simulations, we plot the ratio of ED and $w_{\delta +}$  in
  Illustris to those of MBII in Figure~\ref{F:fig1_edwdpratio}. Given the similarity of the radial
  scaling in the highest mass bin for ED and both the mass bins for $w_{\delta +}$, it is possible to fit the ratio of correlation functions in the range $0.1 - 10 \hmpc$ to
  a straight line. In the mass bin, $10^{9.0-10.5}\hMsun$, we can see from the figure that the radial scaling of ED is not similar between the two simulations and hence, we do not fit the ratio to a straight line in this mass bin. We find that the ED correlation in Illustris is smaller by a factor of $\sim 2.4$
  in the stellar mass bin $10^{10.5-12.0}\hMsun$. Similarly the $w_{\delta +}$ correlation function
  in Illustris is smaller by a factor of $\sim$ $2.8$ and $3.1$ in the lower and higher mass bins
  respectively.

In addition to the differences in galaxy shapes and alignments, the effects due to box
size for MBII and Illustris should be considered as a possible cause of differences in two-point
correlation function amplitude. Since the Illustris
simulation has a smaller volume, the correlation function is suppressed due to the
absence of large scale modes \citep[e.g.,][]{{2005MNRAS.358.1076B},{2006MNRAS.370..691P}} and the
dark matter correlation function is smaller in Illustris by as much as $\sim 20 \%$, with some scale dependence. In order to take
this into account, we compute the ratio of the correlation functions of the shapes of the stellar
matter with the correlation functions of the shapes  of dark matter within the same simulation. This ratio should essentially divide out
such box size effects.  

In the bottom left panel of Figure~\ref{F:fig1_edwdp}, we compare this ratio
for the ED correlation, which we denote $b_{\omega \delta}$. On small scales, the differences in the
amplitude of $b_{\omega \delta}$ in MBII and Illustris are relatively smaller than the differences
in the ED correlation itself. However, on large scales, the ED correlation for the shapes of dark matter is similar for MBII and Illustris within the same stellar mass bins, and hence we observe a significantly smaller value of $b_{\omega \delta}$ in Illustris. In the bottom right panel, we show the bias, $b_{w \delta +}$ , which is
obtained by normalizing the $w_{\delta +}$ for the shapes of stellar component with that of the
shapes of dark matter component. There are smaller differences in amplitude when comparing this
quantity in Illustris and MBII, especially in the higher stellar mass bin at small scales from $\sim
10^{-2}$ to $1~\hmpc$, but there are significant differences in the lower stellar mass bin at all
scales. Considering all factors together, we conclude that in addition to box size effects,
differences in the distributions of shapes and misalignment angles also lower the amplitude of
density-shape correlation functions in Illustris compared to those for comparable samples in MBII.
         
\section{Morphological Classification in Illustris and MBII}\label{S:morph} 
Here, we present the results of classifying the galaxies in Illustris and MBII into disks and
ellipticals using the kinematic bulge-to-disk decomposition. We then compare the
shape distributions and two-point intrinsic alignments statistics of disks and elliptical galaxies. 

\subsection{Fraction of disk galaxies}\label{SS:diskfrac}
   
\begin{figure}
\begin{center}
\includegraphics[width=3.2in]{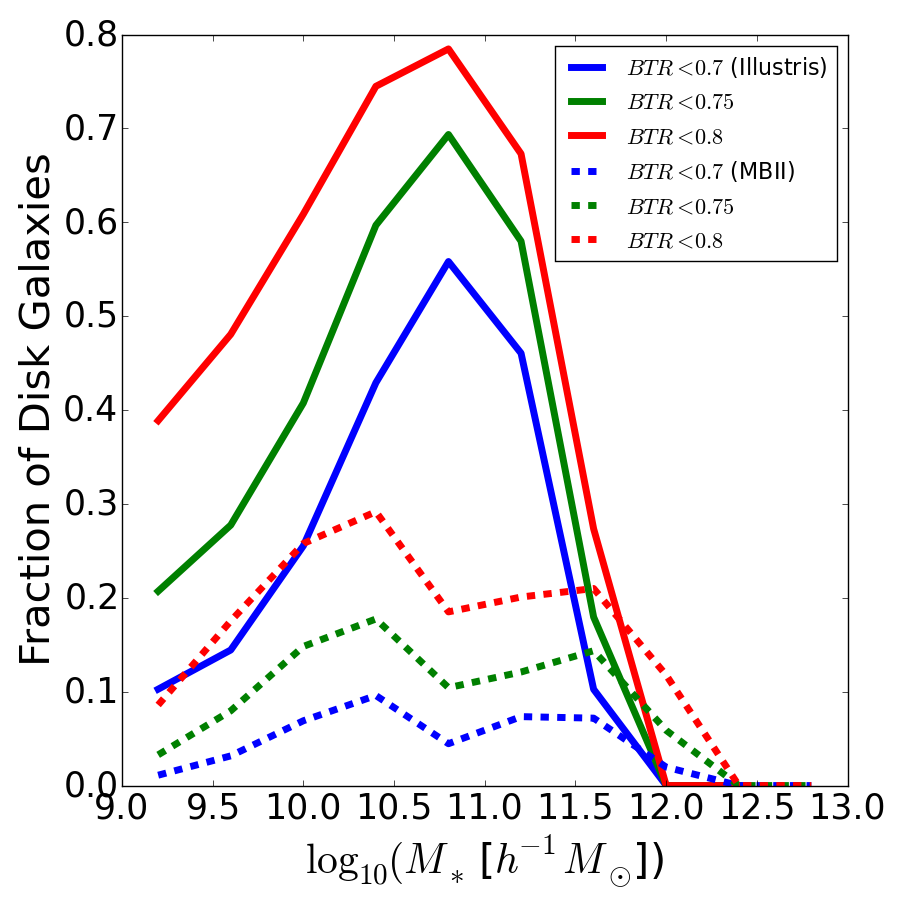}
\caption{\label{F:fig2diskfrac} Fraction of galaxies in MBII and Illustris at $z=0.06$ for different
  thresholds of the bulge-to-total ratio: $BTR$ $<$ $0.7$, $0.75$, $0.8$. Our adopted threshold for
  the rest of this work is that galaxies with $BTR < 0.7$ are classified as disk galaxies.}
\end{center}
\end{figure} 

\begin{figure}
\begin{center}
\includegraphics[width=3.2in]{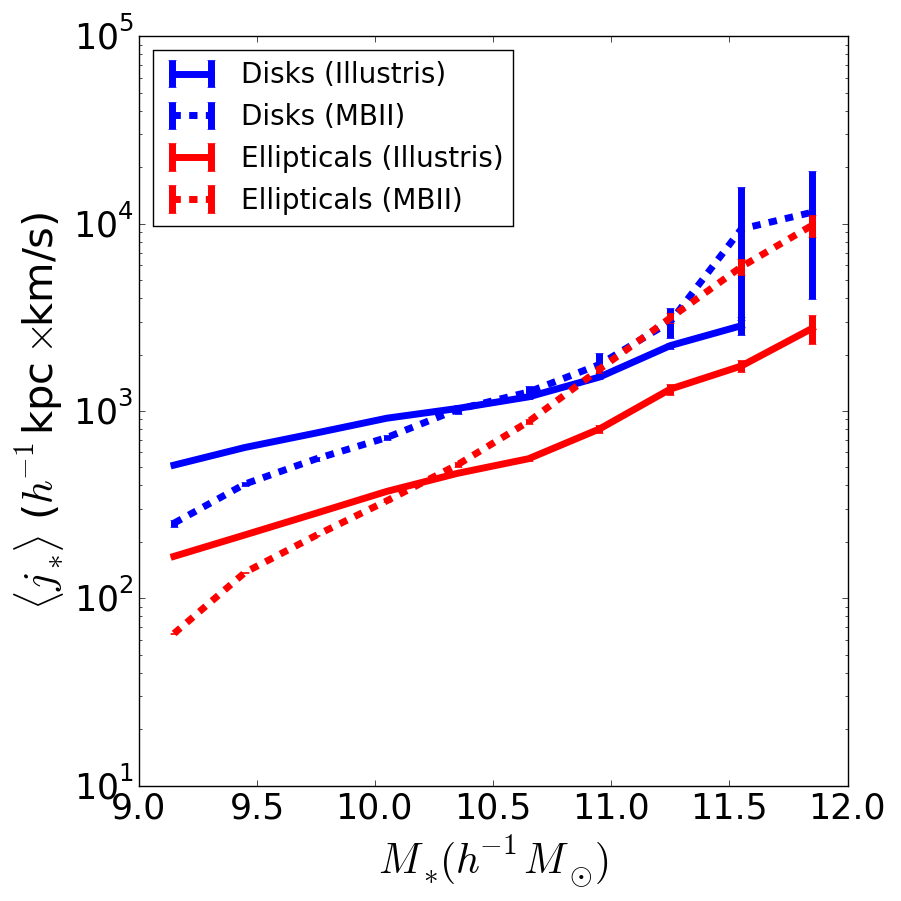}
\caption{\label{F:fig13_jspecam}Mean specific angular momentum of disks and elliptical galaxies
  in MBII and Illustris at $z=0.06$. 
}
\end{center}
\end{figure}

 Using the method described in Section~\ref{S:morphbd}, we calculated the bulge-to-total ratio for each of the simulated galaxies in Illustris and MBII. We note that the threshold adopted to classify galaxies as disks based on the bulge-to-total ratio varies across different studies based on simulations. For instance, \cite{2014MNRAS.444.1518V} classify galaxies with $BTR < 0.7$ as disks while \cite{2009MNRAS.400...43C} adopt a threshold of $BTR < 0.8$. Figure~\ref{F:fig2diskfrac} shows the fraction of galaxies in Illustris and MBII for
different thresholds in the bulge-to-total ratio with $BTR < 0.7, 0.75$, and
$0.8$. 
For our adopted threshold of $0.7$, the fraction of disk galaxies in Illustris varies from $10-50 \%$ in the stellar mass range
$10^{9}$--$10^{12}\hMsun$, while it is below $10\%$ for galaxies in MBII. It rises to $20\%$ with 
a smaller threshold, such as $BTR < 0.8$. We also note that using a smaller threshold on the $BTR$
can lead to some differences in the distributions of axis ratios and misalignment angles. However,
the changes in the two-point statistics are not significant. Hence, in the rest of this paper, we
only show results of morphological classification such that galaxies with $BTR < 0.7$ are classified as disk galaxies and the rest as ellipticals.

\begin{figure*}
\begin{center}
\includegraphics[width=3.2in]{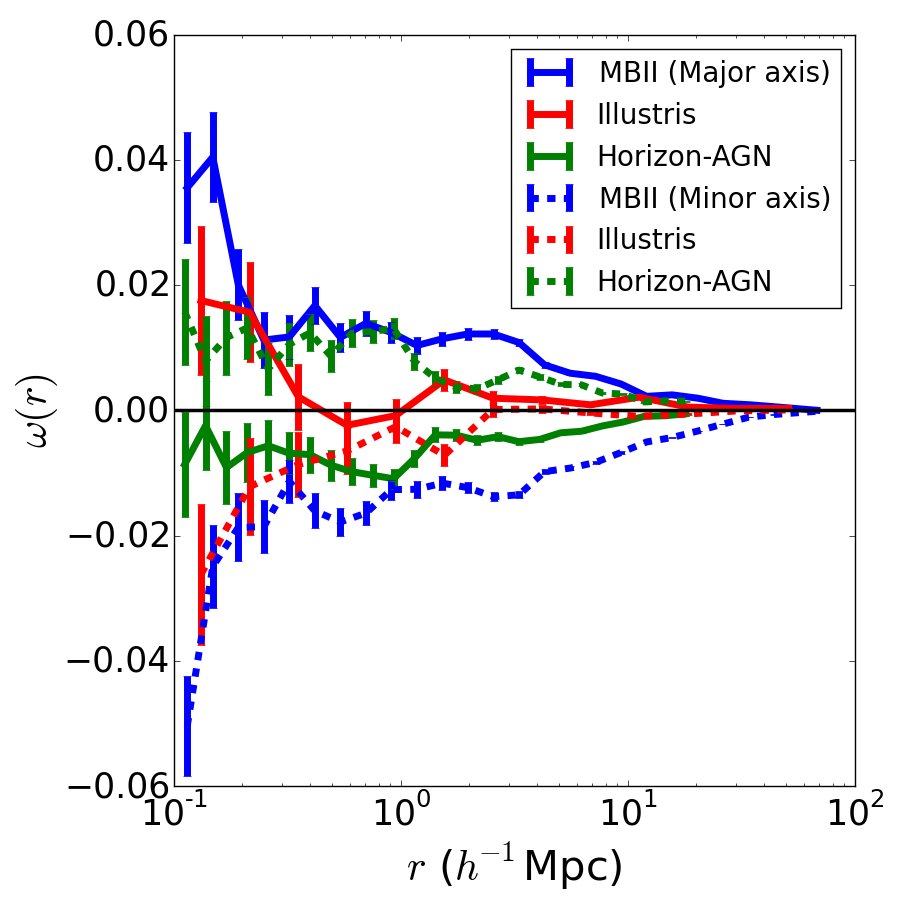}
\includegraphics[width=3.2in]{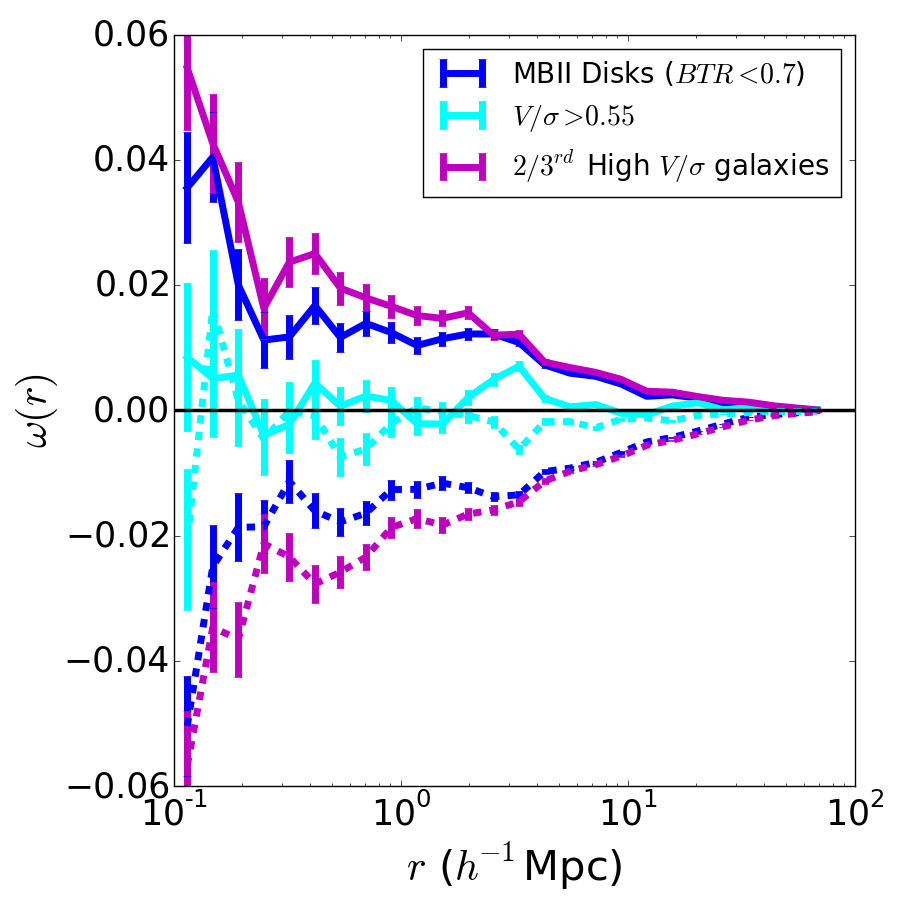}
\caption{\label{F:fig14ed_db} Comparison of the ED correlation of the orientation of disk galaxies
  with the location of ellipticals in MBII, Illustris, and Horizon-AGN simulations
  \protect\citep{2015arXiv150707843C}. In the right panel, the ED correlation in MBII is compared
  using various definitions of the disk galaxy sample. In both the panels, the solid lines represent the ED correlation of the disk major axes while the dashed lines represent the correlation of the disk minor axes.}
\end{center}
\end{figure*}

To further understand the differences between the properties of disks and elliptical galaxies, we
compare their mean specific angular momenta as a function of stellar mass in
Figure~\ref{F:fig13_jspecam}. The mean specific angular momenta of disks and elliptical galaxies
were found to be consistent with observations in the Illustris simulation
\citep{2015ApJ...804L..40G}, with disks having a larger specific angular momentum. Here, we observe
that the disk galaxies in MBII also have a larger specific angular momentum than ellipticals at
lower mass. However, at high stellar mass, the mean specific angular momenta of disks and
ellipticals in MBII are very similar.  When we change the threshold on $BTR$ to $0.75$ or $0.8$, the
specific angular momentum of disk galaxies decreases $\sim 10-20 \%$ for subhalos of stellar mass
below $\sim 10^{10.5}\hMsun$ and increases by a similar amount for subhalos of higher stellar mass. 

Comparing the specific angular momentum in MBII and Illustris for fixed galaxy type, we see that it
is smaller in MBII at low stellar masses, but higher in MBII at higher stellar mass. It has been
shown in \cite{2015ApJ...804L..40G} that the radio mode decreases the specific angular momentum by
$\sim 20-50 \%$.  Thus, the difference at high stellar mass 
may be due to the absence of radio-mode in the MBII simulation, and also an increase in the number
of baryonic particles at higher stellar mass, which can account for the numerical resolution effects
which reduce angular momentum in SPH simulations.

We also compared the angle of orientation between the directions of the angular momentum of stellar component with the angular momentum of dark matter component of the subhalo. We found that in both MBII and Illustris, the
 alignment of the angular momentum of star particles with that of dark matter particles is larger in
 disk galaxies than ellipticals. This is consistent with the findings of
 \cite{2015arXiv150303501T}, who analyzed disks and elliptical galaxies in the Magneticum Pathfinder
 Simulations. Recently, \cite{2015arXiv150707843C} analyzed the 3D orientations of disk galaxies in
 the Horizon-AGN simulation with respect to the location of elliptical galaxies at $z=0.5$, and
 found that the orientation of the disk major axis is anti-correlated with the location of
 ellipticals. We made a similar analysis at $z=0.6$ using the disk galaxies in MBII and Illustris
 with our adopted disk classification. In the left panel of Figure~\ref{F:fig14ed_db}, we plot the
 ED correlation of the orientation of disk major and minor axis with respect to the location of
 ellipticals as a function of separation. Comparing with the results of \cite{2015arXiv150707843C}
 as shown on the plot, we observe that unlike the disks in Horizon-AGN, the major axes of disk
 galaxies in both MBII and Illustris are positively correlated with the location of ellipticals. The minor axes of the disk galaxies are tangentially oriented towards the direction of ellipticals. The direction of the spin or angular momentum of the stellar component of the galaxy is more aligned with the minor axis when compared with the major axes. So, similar to the disk minor axes, the angular momentum of disks is tangentially aligned with respect to ellipticals.  

Note that \cite{2015arXiv150707843C} used the ratio of mean azimuthal velocity of stars to their
velocity dispersion, $V/\sigma$ and classified all galaxies with $V/\sigma > 0.55$ as disks. This
threshold is chosen such that $2/3$ of their galaxy sample is classified as disks. We verified that
our results do not change sign when adopting a different morphological classifies for disk
galaxies. In the right panel of Figure~\ref{F:fig14ed_db}, we compare the results in MBII using the
galaxies classified as disks using $BTR < 0.7$ (our adopted selection throughout this work) with two
other disk galaxy selection criteria: the galaxies for which $V/\sigma > 0.55$, and $2/3$ of the
sample with the highest $V/\sigma$. The results are qualitatively similar with these different
definitions, so the difference in sign observed in Horizon-AGN compared with MBII and Illustris is
unlikely to arise from differences in morphological classifiers. We conclude that differences in
properties of disk galaxies in SPH and AMR simulations and also for different choices of sub-grid physics using the same hydrodynamical code should be explored further.

\subsection{Shapes and misalignment angles of disks and elliptical galaxies}
\begin{figure*}
\begin{center}
\includegraphics[width=3.2in]{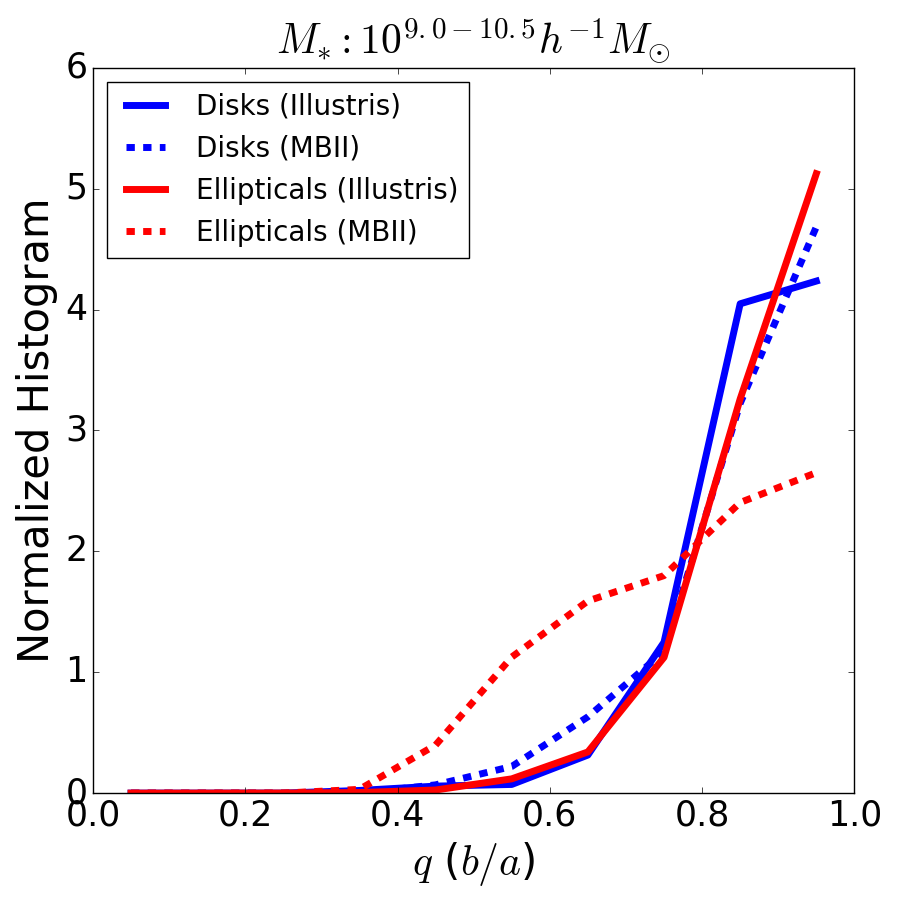}
\includegraphics[width=3.2in]{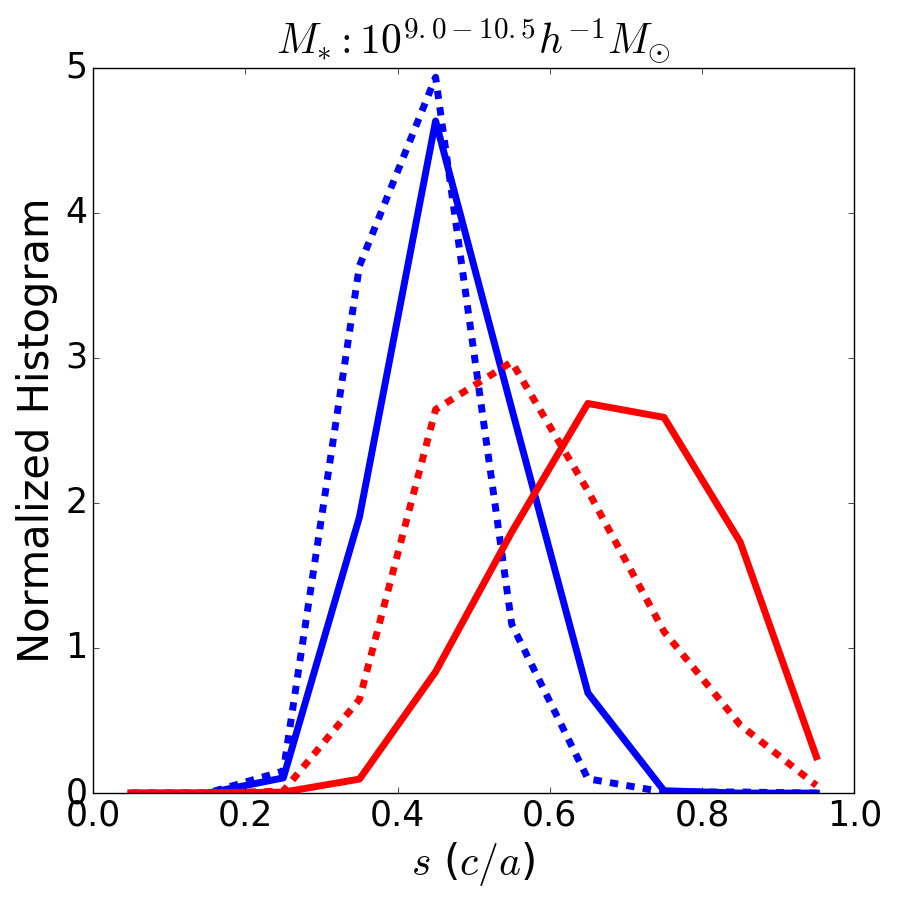}
\caption{\label{F:fig8_qs} Normalized histogram of the axis ratios (left: $q$, right: $s$) of 3D shapes of elliptical and disk galaxies in MBII and Illustris in the stellar mass bin $10^{9}$--$10^{10.5}h^{-1}M_{\odot}$.}
\end{center}
\end{figure*}

\begin{figure}
\begin{center}
\includegraphics[width=3.2in]{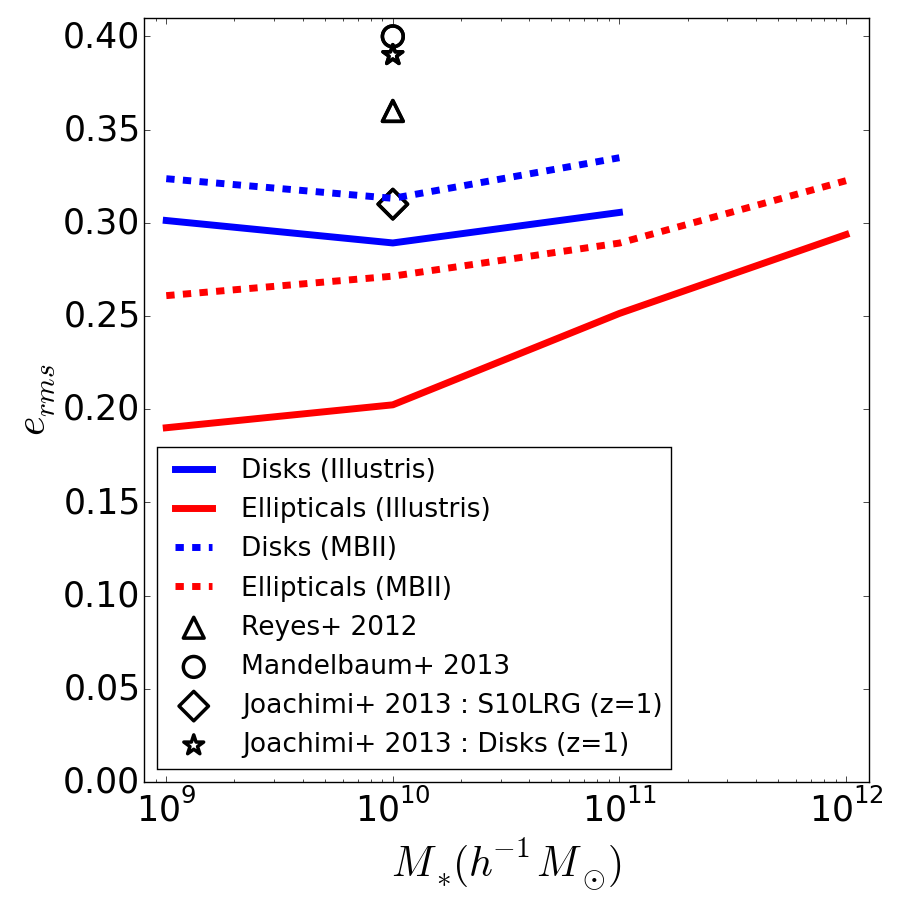}
\caption{\label{F:fig8_erms} RMS ellipticities of the projected shapes of elliptical and disk
  galaxies in MBII and Illustris based on thresholds in stellar mass.
}
\end{center}
\end{figure}

\begin{figure*}
\begin{center}
\includegraphics[width=3.2in]{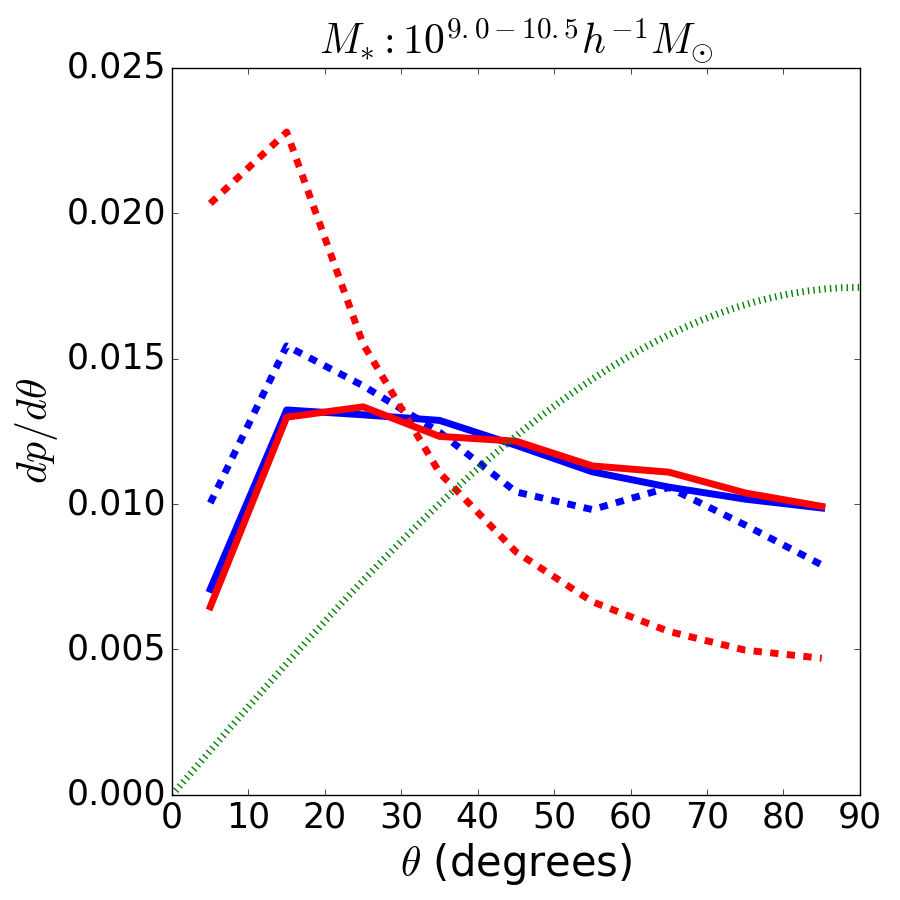}
\includegraphics[width=3.2in]{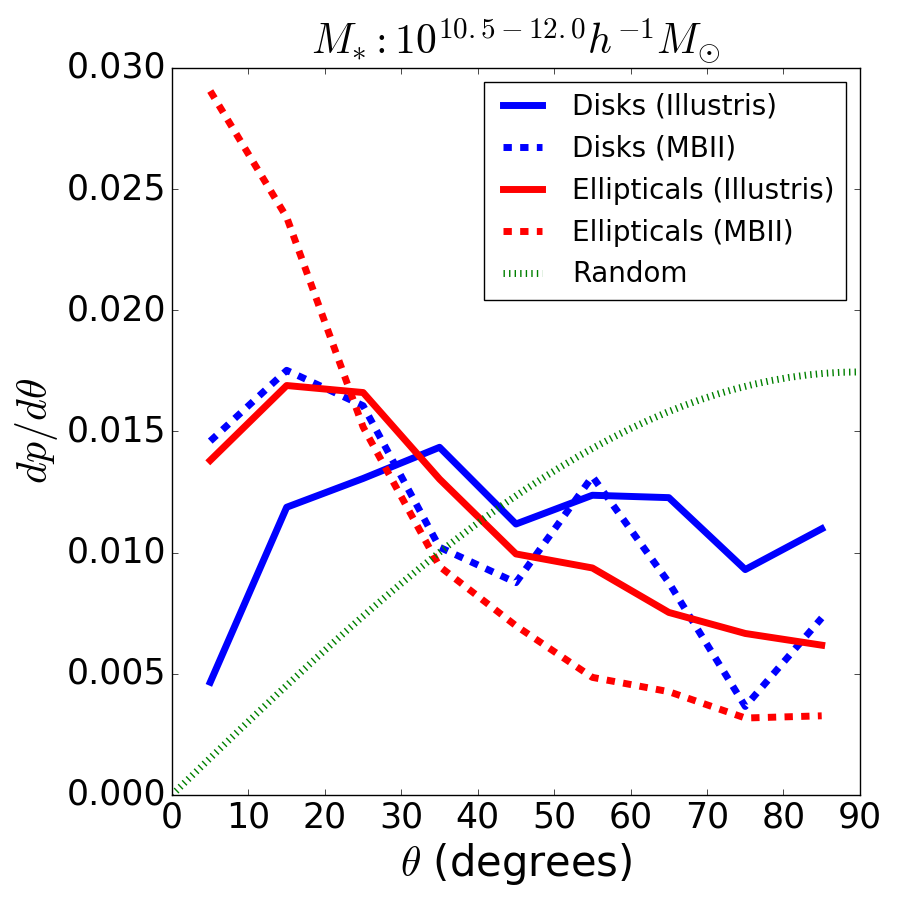}
\caption{\label{F:fig9_ma}Normalized histogram of the misalignment angles of 3D shapes of elliptical
  and disk galaxies in MBII and Illustris within two stellar mass bins: $10^{9-10.5}h^{-1}M_{\odot}$
  (left) and $10^{10.5-12}h^{-1}M_{\odot}$ (right).
}
\end{center}
\end{figure*}

\begin{table}
\begin{center}
\caption{\label{T:tab1a} Mean axis ratios, $\langle q \rangle$ and $\langle s \rangle$, of disks and elliptical galaxies
  in Illustris and MBII. 
}
\begin{tabular}{@{}lcccc}
\hline
 & \multicolumn{2}{c}{Illustris} & \multicolumn{2}{c}{MBII}\\
\hline
 $M_*$ (\hMsun) & Disks & Ellipticals & Disks & Ellipticals \\
\hline
\multicolumn{5}{c}{$\langle q\rangle$}\\
$10^{9}-10^{10.5}$ & $0.87$ & $0.88$ & $0.86$ & $0.77$\\
$10^{10.5}-10^{12}$  & $0.90$ & $0.84$ & $0.91$ & $0.75$\\
\hline
\multicolumn{5}{c}{$\langle s\rangle$}\\
$10^{9}-10^{10.5}$ & $0.47$ & $0.68$ & $0.42$ & $0.57$\\
$10^{10.5}-10^{12}$ & $0.47 $ & $0.61$ & $0.43$ & $0.53$\\

\end{tabular}
\end{center}
\end{table}

\begin{table*}
\begin{center}
\caption{\label{T:tab3a} Mean misalignment angles in 3D, $\langle \theta \rangle$ (degrees), of disks and elliptical galaxies in Illustris and MBII.}
\begin{tabular}{@{}lcccc}
\hline
 $M_*$ (\hMsun) & \multicolumn{2}{c}{Illustris} & \multicolumn{2}{c}{MBII}\\
\hline
  & Disks & Ellipticals & Disks & Ellipticals \\
\hline
$10^{9}-10^{10.5}$ & $44.61 \pm 0.40^{\circ}$ & $45.13 \pm 0.18^{\circ}$ & $41.42 \pm 0.68^{\circ}$ & $31.01 \pm 0.11^{\circ}$\\
$10^{10.5}-10^{12}$ & $46.46 \pm 0.74^{\circ}$ & $36.68 \pm 0.75^{\circ}$ & $36.85 \pm 2.07^{\circ}$ & $25.85 \pm 0.47^{\circ}$\\
\end{tabular}
\end{center}
\end{table*}

We compare the shapes and misalignment angles of the kinematically-classified disks and elliptical
galaxies in Illustris and MBII. In Figure~\ref{F:fig8_qs}, we plot the normalized histograms of the
axis ratios, $q~(b/a)$ and $s~(c/a)$ for galaxies with stellar mass in the range $10^{9}$--$10^{10.5}h^{-1}M_{\odot}$. 
Disk galaxies have larger values
of $q$ and smaller values of $s$ than elliptical galaxies in both Illustris and MBII. This reflects
the fact that disk galaxies have a more oblate shape than elliptical galaxies. 
Comparing the axis ratios in Illustris and MBII, we find that elliptical
galaxies are rounder in Illustris, consistent with the earlier results when considering mass dependence
alone (Sec.~\ref{SS:shapes-mass}). 

However, the distributions for disk galaxies show that disks in MBII have slightly larger
values of $q$ but smaller values of $s$ than in Illustris. This implies that disk galaxies in MBII
are thinner (more oblate) 
compared to those in Illustris. Galaxies in a higher stellar mass bin,
$10^{10.5}$--$10^{12}\hMsun$ (not shown), follow similar trends with respect to axis ratios. 
The mean axis
ratios, $\langle q \rangle$ and $\langle s \rangle$, for disks and ellipticals in Illustris and MBII for
the two stellar mass bins are given in Table~\ref{T:tab1a}. Similarly, in the case of projected
shapes, the RMS ellipticities for disk galaxies in  MBII are larger. 
The RMS ellipticities based on a stellar mass threshold are
shown in Figure~\ref{F:fig8_erms} and compared with observations. Note that when compared to
  observational measurements, the RMS ellipticities are smaller even for disk galaxies in
  MBII. However, a direct quantitative comparison with observations is difficult due to different
  methods adopted to measure the observed and simulated galaxy shapes. A detailed discussion on this
  comparison can be found in \cite{2014MNRAS.441..470T}.

In order to understand the orientation of disks and ellipticals with the shape of their host dark
matter subhalos, we compare the histograms of misalignment angles in Figure~\ref{F:fig9_ma}. In
general, disk galaxies are more misaligned with their host dark matter shapes when compared with
elliptical galaxies. The mean misalignment angles are provided in Table~\ref{T:tab3a}. Note that
disks and ellipticals in the lower mass bin of Illustris, $10^{9}$--$10^{10.5}\hMsun$, have similar histograms of
misalignment angles. This may be a mass-dependent effect where misalignments increase as we go
to lower masses. If we increase the lower mass threshold of the bin, the histograms are shifted such
that disks tend to be more misaligned than ellipticals. 

Based on the results shown in this section, we find that the disk and elliptical galaxies in
  the MBII and Illustris simulations have qualitatively similar shapes and misalignment angle
  distributions. However, there are differences in the disk fraction and the amplitude of
  misalignments in the galaxies of the two simulations. This is likely due to the differences in
  sub-grid physics in the two simulations. A detailed study considering the effects of various
  baryonic feedback implementations in simulations on the galaxy alignments is deferred for future
  study.

\subsection{Two-point intrinsic alignment statistics of disks and elliptical galaxies}
Using the measured shapes of disks and elliptical galaxies, we compare the two-point shape
correlation functions of disk galaxies with those of ellipticals in both simulations.

\begin{figure*}
\begin{center}
\includegraphics[width=3.2in]{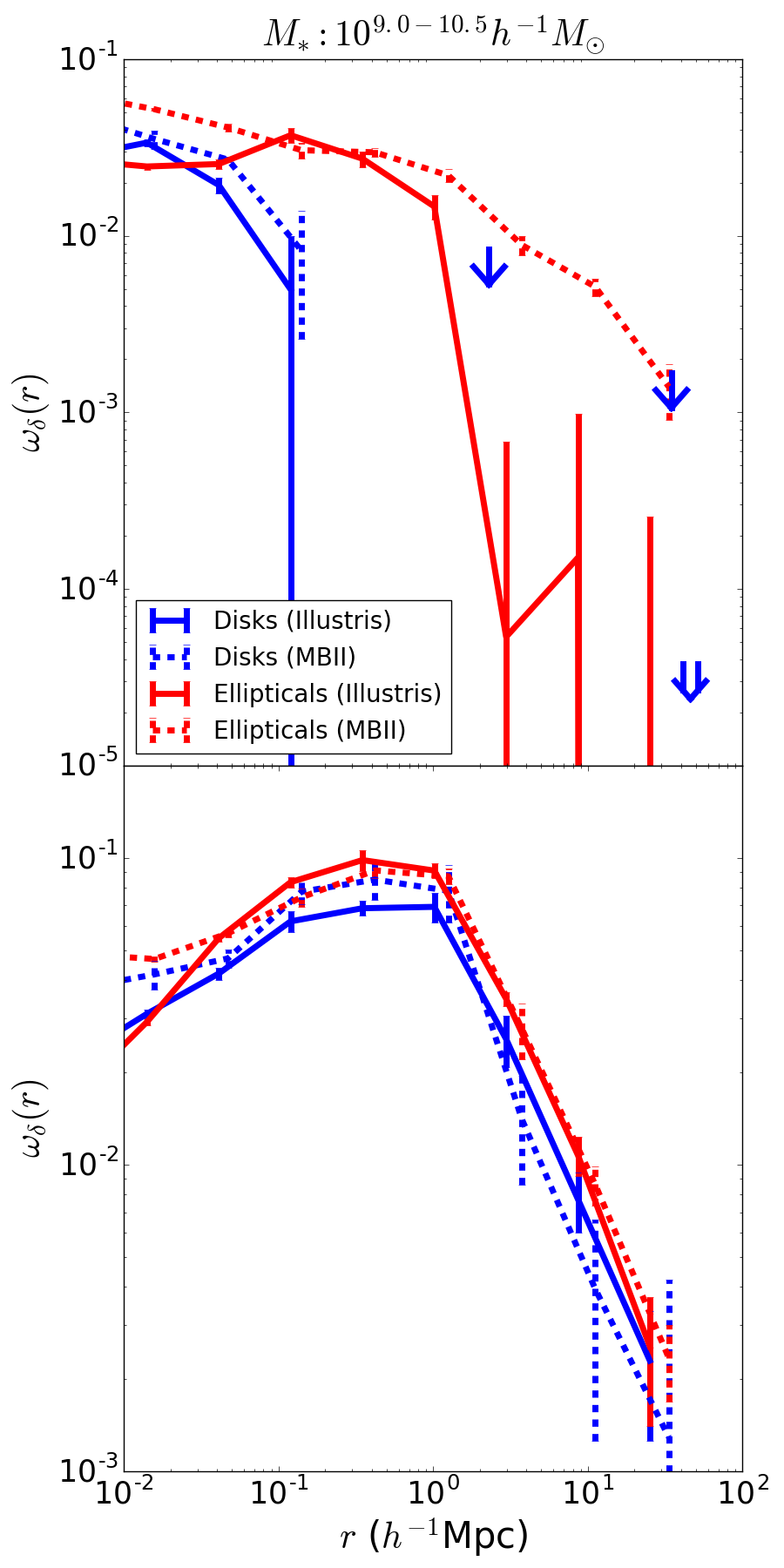}
\includegraphics[width=3.2in]{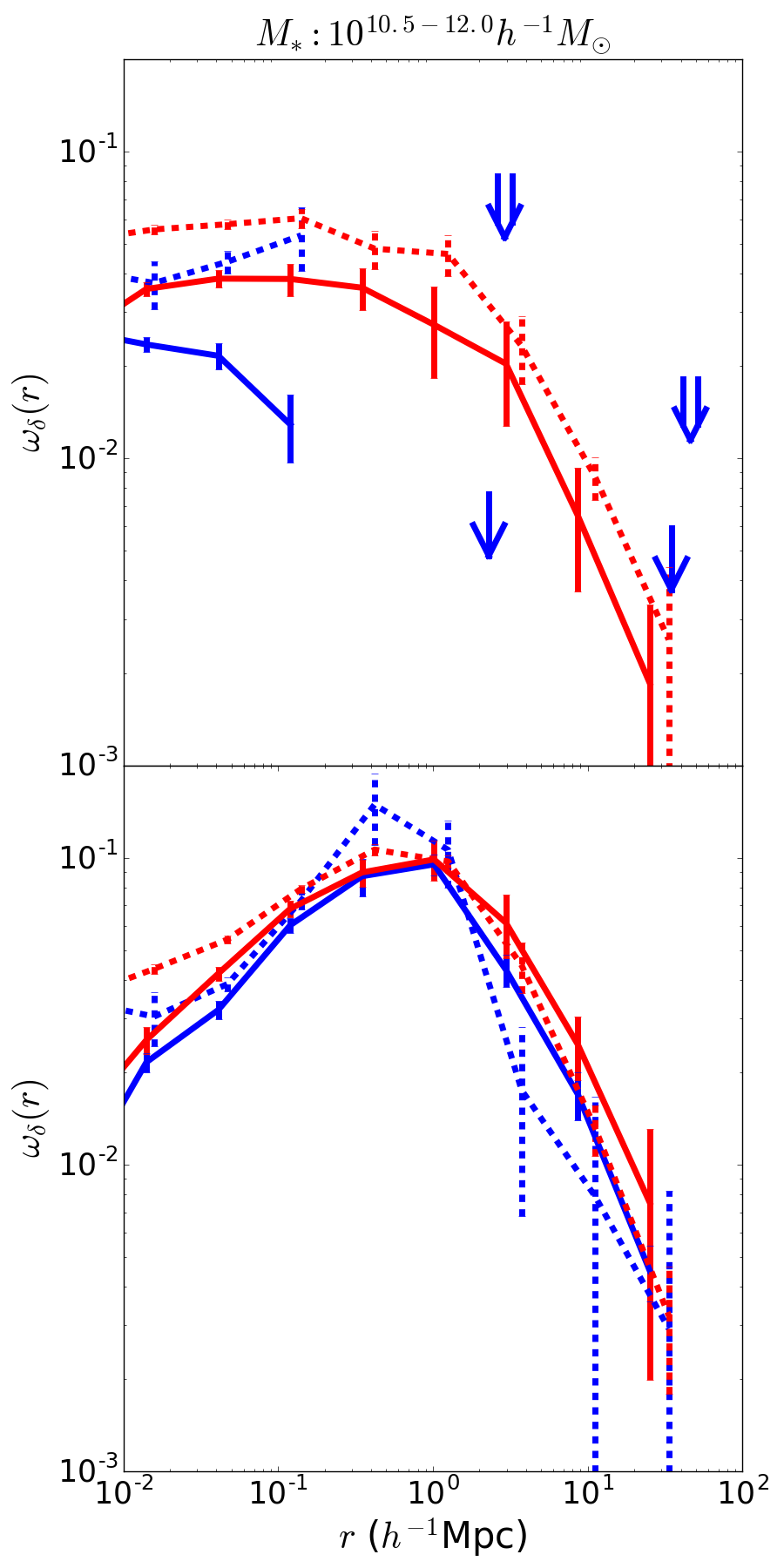}
\caption{\label{F:fig2_ed} ED correlation functions of the shapes of the stellar component ({\em top} panel) and the dark matter component ({\em bottom} panel) of elliptical and disk galaxies
  galaxies in MBII and Illustris within two stellar mass bins: $10^{9}$--$10^{10.5}h^{-1}M_{\odot}$ (left)
  and
  $10^{10.5}$--$10^{12}h^{-1}M_{\odot}$ (right). In the {\em top} panel, we only show the $1\sigma$ upper limits of the ED correlation for disk galaxies on scales above $0.1\hmpc$, represented by the {\em bottom (tip)} of 
arrows pointing downward in Illustris ($\downarrow$) and MBII ($\Downarrow$). Note the $y$-axis
limits are different in the top left panel compared to the other panels, which should be taken into account while comparing the figures.  
}
\end{center}
\end{figure*}

\begin{figure*}
\begin{center}
\includegraphics[width=3.2in]{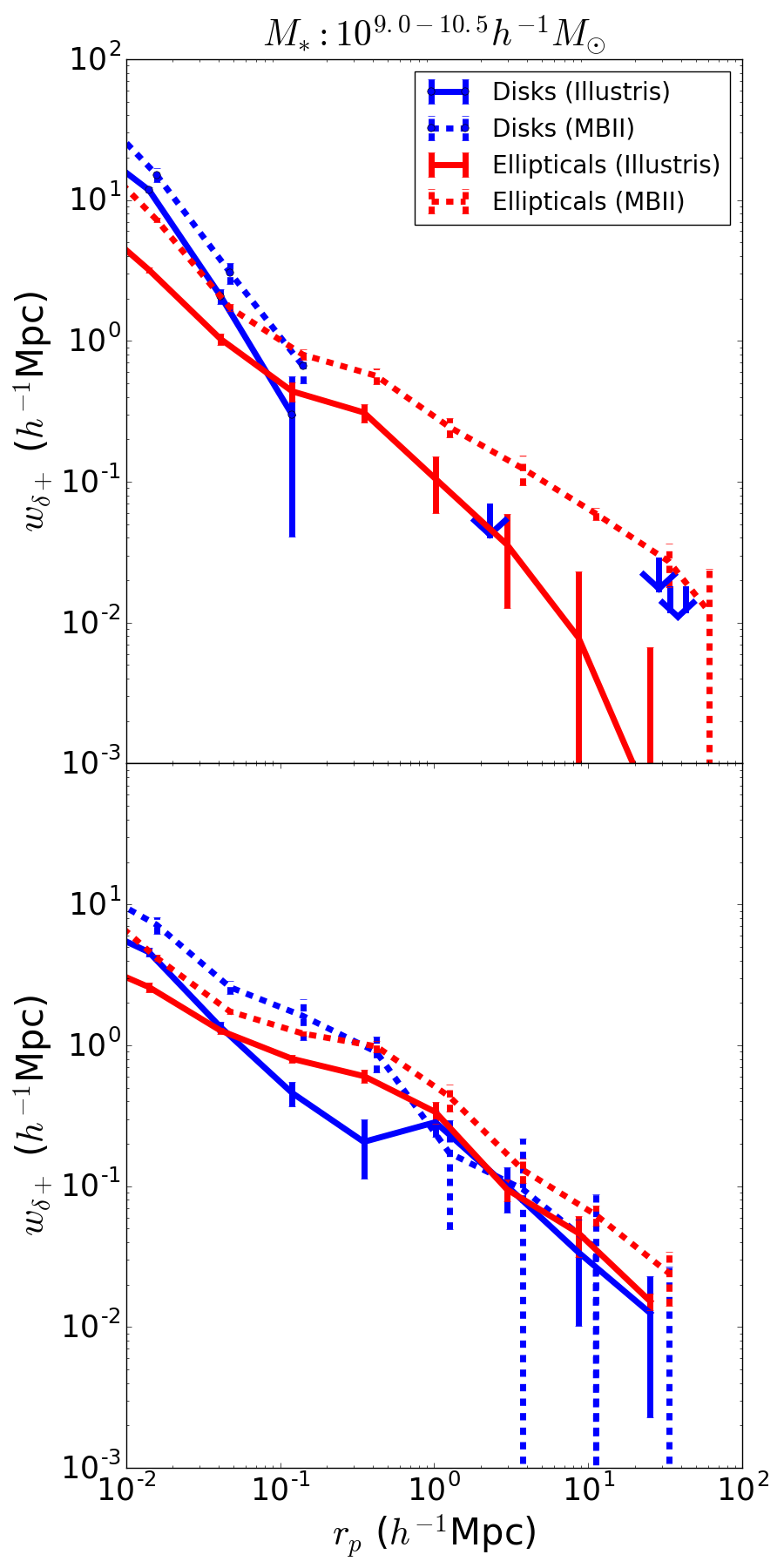}
\includegraphics[width=3.2in]{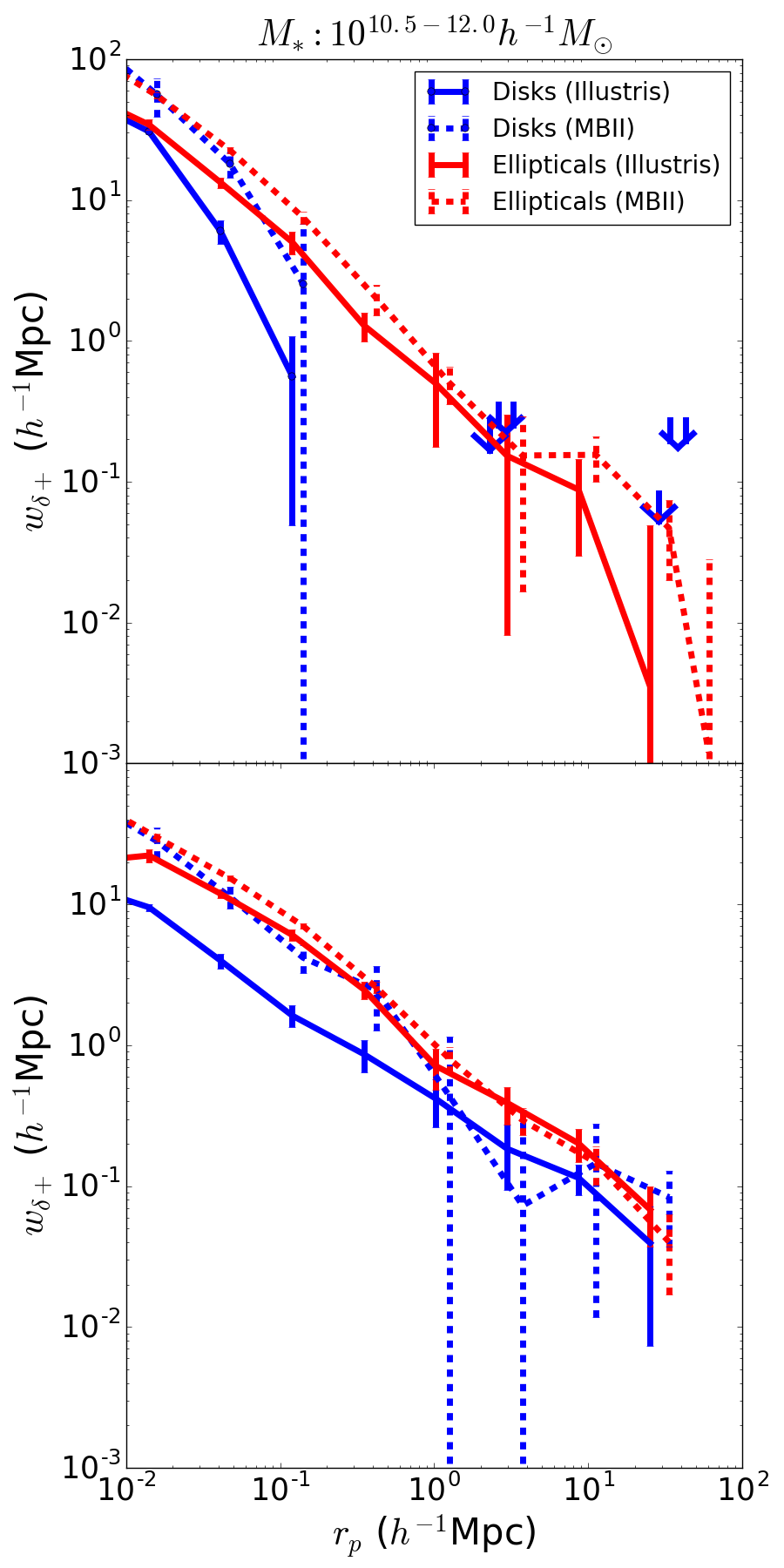}
\caption{\label{F:fig2_wdp} $w_{\delta +}$ correlation functions of the shapes of the stellar component ({\em top} panel) and the dark matter component ({\em bottom} panel) of elliptical and
  disk galaxies in MBII and Illustris within two stellar mass bins: $10^{9}$--$10^{10.5}h^{-1}M_{\odot}$
  (left) and 
  $10^{10.5}$--$10^{12}h^{-1}M_{\odot}$ (right). On scales above $0.1\hmpc$, we only show the $1\sigma$ upper limits of the $w_{\delta +}$ signal for disk galaxies, represented by the {\em bottom (tip)} of 
arrows pointing downward in Illustris ($\downarrow$) and MBII ($\Downarrow$). 
}
\end{center}
\end{figure*}

Figure~\ref{F:fig2_ed} shows the ED correlation function in two stellar
mass bins: $10^{9}$--$10^{10.5}\hMsun$ and $10^{10.5}$--$10^{12}\hMsun$. As shown, 
the disk galaxies have a smaller ED correlation function than ellipticals; indeed, the correlation
function is consistent with zero for disk galaxies in both simulations and mass bins on scales above $\sim
100h^{-1}$kpc. Hence, we only show the $1\sigma$ upper limits of the signal on these scales, represented by the {\em bottom (tip)} of 
arrows pointing downward in Illustris ($\downarrow$) and MBII ($\Downarrow$). 
The disk vs.\ elliptical difference is due to the
larger misalignment of disk galaxy shapes with their host dark matter subhalo shapes. As discussed further below, we find that the ED correlation is similar for the shapes of dark matter subhalos of disks and ellipticals, while the galaxy misalignment suppresses the correlation. For the same
reason, 
the correlation functions are larger in MBII than in Illustris for all samples. 

 To further understand the differences in the ED correlations of disks and ellipticals, we compare
 the ED correlations of the dark matter subhalos of disks and ellipticals in the bottom panels of
 Figure~\ref{F:fig2_ed}. In both Illustris and MBII, the ED correlation function of the dark matter
 subhalos hosting disk galaxies is significant even on large scales, and has similar radial scaling
 compared with that of ellipticals. The small differences in amplitude might relate to the slightly
 different subhalo mass distributions for disk and elliptical galaxies within these mass bins.  The
 strong similarity between the results for subhalos hosting disk and elliptical galaxies reinforces
 our conclusion 
 that the suppression in the ED correlation for the stellar components of disk galaxies is largely
 due to stronger misalignment with the shape of their host dark matter subhalo. 

In Figure~\ref{F:fig2_wdp}, we compare the projected shape-density ($w_{\delta +}$) correlation
function in the same two stellar mass bins. Similar to the ED correlation, the $w_{\delta +}$ for
disk galaxies is noisy on scales above $0.1 \hmpc$, and only the upper limits of the signal are
shown above these scales. However, on smaller scales, the amplitude of $w_{\delta +}$ for disk
galaxies is larger than that for ellipticals in both simulations, due to the fact that disk galaxies
have larger ellipticities than elliptical galaxies.  

These predictions for disk galaxies can be compared against the measurements of
\cite{2007MNRAS.381.1197H}, which has a null detection of $w_{g+}$ for most of their blue galaxy
samples, but a weak detection of intrinsic alignments for their most luminous blue galaxy sample,
consistent with the amplitude of red galaxy $w_{g+}$ in the same luminosity bin. While the amplitude
of $w_{\delta +}$ is comparable for disks and ellipticals at small scales around $0.1\hmpc$, we do not detect the
signal for disks at scales around $1-10\hmpc$, where the measurements in \cite{2007MNRAS.381.1197H}
are made. We also note here that \cite{2011MNRAS.410..844M} investigated the intrinsic alignments of
blue galaxies from the WiggleZ sample at $z \sim 0.6$ and find a null detection for $1-10 \hmpc$. However, when considering the stellar mass bin $10^{11}$--$10^{12}h^{-1}M_{\odot}$ (not
shown), the $w_{\delta +}$ for disk galaxies in MBII is higher than for both samples shown in this
section, and is comparable in magnitude with the amplitude of the weak signal detected for blue
galaxies in \cite{2007MNRAS.381.1197H} at scales around $\sim 1\hmpc$. 

\section{Conclusions}\label{S:conclusions}     
In this paper, we studied the shapes and intrinsic alignments of disk and elliptical galaxies in the
MassiveBlack-II and Illustris simulations. 
The galaxy stellar mass function is similar in both the simulations at high mass
range, while at lower stellar masses, MBII has a higher number density of galaxies. We restrict our
analysis to stellar masses ranging from $10^{9}$--$10^{12}\hMsun$, for which the galaxies have a
minimum of 1000 star particles.

We first compared the galaxy shapes and alignments in Illustris and MBII based on stellar mass
alone, without considering disks and ellipticals separately. While the two simulations show similar
trends in galaxy shapes with stellar mass (rounder at lower stellar mass), the galaxy shape
distributions are rounder in Illustris than in MBII at fixed mass. By measuring the
orientation of the shape of the stellar component with respect to the major axis of the host dark
matter subhalo, we find that both simulations show similar trends with mass, with stronger
misalignments at lower mass.  However, at fixed mass, galaxies are more misaligned with their host subhalo shapes in the
Illustris simulation than in MBII. 

 Due to the larger
misalignment of the galaxy stellar components with the density field, the ellipticity-direction (ED)
correlation function has a smaller amplitude in Illustris at fixed stellar
mass. At around $1\hmpc$, the correlation function is larger in MBII by a factor of $\sim 1.5-3.5$. 

Similarly, the
amplitude of the projected shape-density correlation function, $w_{\delta +}$, is smaller in
Illustris by a factor of $\sim 1.5-2.0$ at transverse separation of $1\hmpc$ due to both the larger misalignments and the smaller ellipticities. These
differences in the amplitudes of the intrinsic alignment correlation functions are significant even after accounting for the bias
in the dark matter correlations due to the smaller volume in Illustris. 
We further find that the
mass- and scale-dependence of the $w_{\delta +}$ two-point statistic is similar in Illustris and
MBII, in spite of the different implementations of hydrodynamics and baryonic physics. However, the scale dependence is significantly different in the ED correlation of low mass galaxies. This can be due to the different implementations of hydrodynamic or differences in baryonic feedback models.
We find signs of different physics behind intrinsic
alignments of disk galaxies in MBII and Illustris compared to findings from the Horizon-AGN
simulation \citep{2015arXiv150707843C}, which suggests possible differences in SPH vs.\ AMR
simulations that warrant further investigation. Based on our findings, we conclude that hydrodynamic simulations are a promising tool to study intrinsic alignments. For higher mass galaxies, our results suggest that hydrodynamic simulations can be used to generate templates for the
scale-dependence of intrinsic alignment two-point correlations for use by upcoming surveys that must
remove this effect from weak lensing measurements, provided that the amplitude of the effect is
marginalized over (given observational priors). However, further study on understanding the differences in various simulations is needed to confirm the validity of this conclusion at lower mass and to confirm that it applies with greater statistical precision at high mass.

Galaxies in MBII and Illustris are classified into disks and ellipticals by a dynamical bulge-disk
decomposition method following the procedure adopted in \cite{2009MNRAS.396..696S}, resulting in a
larger fraction of disk galaxies in Illustris than in MBII at fixed stellar mass. The disk
galaxies in both simulations are more oblate than the elliptical galaxies. However,
the disk galaxies in MBII are more oblate than those in Illustris.

Comparing the alignments of the disk galaxies with their host dark matter subhalos, we find that
disk galaxies are more misaligned than ellipticals in both MBII and Illustris by $\sim 20-30 \%$ on
average. Due to this larger misalignment, the disks have a smaller amplitude of ED correlation when
compared with ellipticals (and compared to the ED correlations of their host dark matter subhalo
shapes). Indeed, this correlation function is consistent with zero for the disk 
samples (within our errorbars) above $\sim 100h^{-1}$kpc.  However, the disk galaxies also have
larger ellipticities, which increases the $w_{\delta +}$ correlation on the small scales where it is
detected. Thus, the amplitude of $w_{\delta +}$ for disks is comparable with that of ellipticals at
the same mass  for scales below $0.1\hmpc$, while on large scales it is consistent with a null
detection.  While exploration with larger-volume simulations that have more disks and hence lower
statistical errors is warranted, our results currently support the commonly-made assumption
\citep[e.g.,][]{2015arXiv150608730K} that 
large-scale intrinsic alignments for early-type galaxies are stronger than those for late-type
galaxies.  This finding bodes well for future weak lensing surveys that will be dominated by
galaxies at $z\gtrsim 0.6$, where the disk galaxy fraction is larger than it is at later times.

\section*{Acknowledgments}

AT and RM acknowledge the support of NASA ROSES 12-EUCLID12-0004. AT thanks Shy Genel for discussions regarding morphological classification in Illustris. We also thank Elisa Chisari for providing the data to compare with results from Horizon-AGN simulation.

\bibliographystyle{mnras} \bibliography{draft4f}
\end{document}